\documentclass[12pt,onecolumn,notitlepage,reqno,fleqn,tbtags]{amsart}
\usepackage{amssymb,latexsym}
\begin{document}

 
Two-Center Integrals for $r_{ij} ^{n}$ Polynomial Correlated Wave Functions\\ 
\begin{center}E. V. Rothstein \end{center} 
\begin{center}evrothstein@gmail.com\end{center} 

      All integrals needed to evaluate the wave function of the form
\[
\Psi_{tot} = \tilde{\mathcal{A}} \quad  \left\{ \left[ \prod_j \sum_s a_{s_j} \phi_{s}(j) \right] \left[ 1 + \sum_{j \quad <} \sum_i w_{ij} r_{ij}^{n}   \right] \right\}     
\]
for n = 1 and the Hamiltonian given are combined herein.  For this form of the wave function, the integrals needed can be expressed as a product of integrals involving at most four electrons.  An indication of how to increase or decrease exponents of $r_{ij} ^ {n}$ in steps of one or two is given.  Some indication of how to proceed if the Hamiltonian contains $1 / r_{ij} ^ {3}$  terms  or  if the wave function is of the form   
\[
\Psi_{tot} = \tilde{\mathcal{A}} \quad  \left\{ \left[ \prod_j \sum_s a_{s_j} \phi_{s}(j) \right] \left[ 1 + \sum_{j \quad <} \sum_{l \quad <} \sum_{k} w_{jlk} r_{jl}^{n} r_{lk}^{\nu}   \right] \right\}     
\]
is given.  Consideration of all possible types of integrals 
involving $r_{ij}^{a} r_{kl}^{b} r_{mn}^{c}$ with $a < 0$, $b > 0$, $c > 0$ ; $\left| a \right| = \left| b \right| = \left| c \right| = 1$, is given.  Integrals
are given in analytical form.  These can be evaluated by numerical integration routines. \\
\begin{center}INTRODUCTION\end{center} 
     Much success in ab initio  calculations has been achieved with the use of correlated wave functions (wave functions that included the distance between two electrons explicitly).  For the He $\text{atom}^1$, the Li $\text{atom}^2$, and the $\text{H}_{2}$ $\text{molecule}^3$, these wave functions have yielded the most accurate energy levels and molecular properties.
   Constructing the total wave function as a Slater determinant or antisymmetrized product is tantamount to using the Pauli principle, which excludes two electrons with identical quantum numbers and spin from occupying the same volume element at the same time.  It does not tell us anything about two electrons with opposite spin, which we would expect to repel each other electrostatically.  By including terms dependent upon the interelectronic separation, we cause the probability, calculated from this wave function, of finding two electrons at specified regions of space to decrease when the two electrons approach one another.
    A correlated wave function can be an eigenfunction of spin and angular momentum.  If $\tilde{\mathcal{A}}$(F) is an eigenfunction of the total and z component of spin and angular momentum, then $\Psi$ is an eigenfunction of the $\text{same}^4$.  Using the Nth-order permutation $\text{group}^5$, all spin $\text{states}^6$ for the N electron system can be included:
\[ \Psi = \tilde{\mathcal{A}}(FR) \] ,
\[ F = F(1, 2, 3, \ldots , N) = (N !)^{-1/2} \Phi_{1}(1)  \Phi_{2}(2)   \cdots \Phi_{N}(N)  \] ,   
\[ R = 1 + \sum_{ j \quad < } \sum_l (w_{jl} r_{jl}   +   x_{jl} r_{jl} ^{2} +  y_{jl} r_{jl} ^{3} + \cdots ) \] 
and  $\tilde{\mathcal{A}}$ is the antisymmetrization operator.  
   H is the Hamiltonian (in the Born-Oppenheimer approximation) in atomic units,$Z_{a}$ and $Z_{b}$ are the nuclear charges, R is the distance between nuclei   and b, $r_{i \lambda}$ is the distance between electron i and nucleus $\lambda$ ,  $r_{ij}$ is the distance between electron i and electron j

\[ 
H = -\frac{1}{2} \sum_{i} \nabla_{i} ^{2} - \sum_{i} \left( \frac{Z_{a}}{r_{ia}}  + \frac{Z_{b}}{r_{ib}} \right) + \sum_{i <} \sum_{j} \frac{1}{r_{ij}}            + \frac{Z_{a} Z_{b}}{R}     \] 
The coordinate $\text{system}^{7}$ used is confocal elliptical.  $\phi_{i}$ is  the out-of-plane angle and $R$ is the distance between nuclei $a$ and $b$.      $d\tau_{i}$ is the volume element and $r_{12}$ the interelectronic distance.  We also have
\begin{align*} \xi_{i} &  = \left( r_{ai} + r_{bi} \right) / R   , &  \eta_{i}  = \left( r_{ai} - r_{bi} \right) / R  \\
& 1 \leq \xi_{i} <  \infty  ,   & -1 \leq \eta_{i} \leq 1  , \\
d\tau_{i} & = \frac{1}{8} R^{ 3 }  \left( \xi_{i}^{2} - \eta_{i}^{2} \right)  d\xi_{i} d\eta_{i} d\phi_{i}  , &  0 \leq \phi_{i} < 2 \pi 
\end{align*}
\begin{multline}
r_{12}^{2} = \frac{1}{4} R^{2} \lbrace   \xi_{1}^{2} + \xi_{2}^{2} +  \eta_{1}^2+ \eta_{2}^2 - 2 -2 \xi_{1} \xi_{2} \eta_{1} \eta_{2}  \\
  - 2 \left[ (\xi_{1}^2 - 1) (\xi_{2}^2 -1) (1 - \eta_{1}^2)(1 - \eta_{2}^2)      \right]^{1/2} \cos ( \phi_{1} - \phi_{2} ) \rbrace    
\end{multline}  
The basis functions $\Phi_{s}(j)$ and the total wave function are represented in Eqs. (2) and (3): 
 
\begin{multline}
\Phi_{s}(j) = \xi_{j}^{p_s} \eta_{j}^{q_s} (\xi_{j}^{2} - 1)^{\gamma_{s}/2} (1 - \eta_{j}^{2})^{\nu_{s}/2} e^{-\alpha_{s} \xi_{j}} e^{\beta_{s} \eta_{j}} e^{im_{s}\phi_{j}} ,\\
s=(p_{s}, q_{s}, \gamma_{s}, \nu_{s}, \alpha_{s}, \beta_{s}, m_{s}) , 
  i =  \sqrt{ - 1}
\end{multline}

\begin{multline}
\Psi_{tot} = \tilde{\mathcal{A}} \quad  \left\{ \left[ \prod_j \sum_s a_{s_j} \phi_{s}(j) \right] \left[ 1 + \sum_{j \quad <} \sum_i w_{ij} r_{ij}^{n}   \right] \right\}     
\end{multline}
For molecules with more than two nuclei, the spherical coordinate system and Gaussian $\text{transforms}^8$ or $\zeta$-function $\text{expansions}^{9}$ can be  used for integral evaluation.
\begin{center} CLASSIFICATION OF INTEGRALS \end{center}
The classification of types of integrals involving 
$r_{ij}^{a}r_{kl}^{b}r_{mn}^c$ can be considered in the notation of picture $\text{writing}^{10}$ graph $\text{theory}^{11}$.  For the form of the wave function given in (3), the integrals needed can be expressed as a product of primitive integrals involving at most four electrons.  All the integrals needed to evaluate this wave function involving $r_{ij}$ to the first power are given.
\begin{center} TWO - ELECTRON INTEGRALS \end{center}
We have
\begin{multline}
\langle r_{12}^{2} \rangle =\langle \Phi_{s}(2) r_{12}^{2} \Phi_{t}(1) \rangle = \int d\tau \Phi_{s}(2) r_{12}^{2} \Phi_{t}(1) \\
= \frac{1}{64} R^{8} \pi ^{2}  
 \int_{1}^{\infty} \int_{1}^{\infty} \int_{-1} ^{1} \int_{-1} ^{1} 
(\xi_{1}^{2} - \eta_{1} ^2) (\xi_{2}^{2} - \eta_{2}^2) d\xi_{1}d\xi_{2}d\eta_{1}d\eta_{2} \\  \times \Phi_{s}(2) \Phi_{t}(1) 
 \, \lbrace \,  [ \,  \xi_{1}^{2} + \xi_{2}^{2} +  \eta_{1}^2+ \eta_{2}^2 - 2 -2 \xi_{1} \xi_{2} \eta_{1} \eta_{2} \,  ] \, \delta(m_{s} ; 0) \delta(m_{t} ; 0)  \\  -  \left[ \, (\xi_{1}^2 - 1) (\xi_{2}^2 -1) (1 - \eta_{1}^2)(1 - \eta_{2}^2)      \right]^{1/2} \, [ \,  \delta (m_{s} - 1 ; 0 ) \delta(m_{t} + 1 ; 0 ) \\ 
 + \delta (m_{s} + 1 ; 0 ) \delta (m_{t} - 1 ; 0 ) \, ] \,  \rbrace      
\end{multline}
The $\text{Neumann}^{12}$ expansion for $1/r_{12}$ in prolate elliptical coordinates is:
\begin{multline}
\frac{1}{r_{12}} = \frac{4}{R} \sum_{l=0}^{\infty} \sum_{m=-l}^{l} (-1)^{m}     
\frac{2l + 1}{2} \left[ \frac{(l - |m|)!}{(l + |m|)!} \right]^{2}               P_{l}^{|m|}(\xi_{1 < 2}) \\
\times Q_{l}^{|m|}(\xi_{2 > 1}) P_{l}^{|m|}(\eta_{1}) P_{l}^{|m|}(\eta_{2}) 
e^{im(\phi_{1}-\phi_{2})}
\end{multline}
$P_{l}^{m}(\xi)$ and $Q_{l}^{m}(\xi)$ are associated Legendre polynomials of the first and second kind in the conmplex $\text{plane}^{13}$.  $\xi_{1 < 2}$ and  $\xi_{2 > 1}$ mean the smaller and the larger, respectively, of $\xi_{1}$ amd   $\xi_{2}$.  The $P_{l}^{m}(\xi)$    and     $Q_{l}^{m}(\xi)$ have the     range [ 1 ,$ \infty$ ].  The $P_{l}^{m}(\eta)$ have the range [ -1 , 1 ].  For  further details, see Refs. 13 - 23.  We have
\begin{multline}
P_{l}^{m}(\xi) = \frac{ (\xi^{2} - 1)^{m/2}}{2^{l}l !} \frac{d^{l+m}}{d\xi^{l+m}}  (\xi^{2} - 1 )^{l} , \\
 P_{l}^{m}(\eta) = \frac{ (1 - \eta^{2})^{m/2}}{2^{l}l!} \frac{d^{l+m}}{d\eta^{l+m}}  (\eta^{2} - 1 )^{l} , \\
Q_{l}^{m}(\xi) = (\xi^{2} - 1)^{m/2} \frac{d^{m}}{d\xi^{m}} Q_{l}(\xi) , \\
Q_{l}(\xi) = \frac{1}{2}  P_{l}(\xi) \ln \left( \frac{\xi + 1 }{\xi - 1 }\right) \\              - \sum_{j = 1}^{ \le (l + 1)/2} \frac{2l - 4j + 3 }{(2j - 1)(l - j + 1)}    P_{l - 2j + 1}(\xi)
\end{multline}
The $ \langle 1 / r_{12} \rangle $  ,Eq. (10) can be expressed more concisely,  using the definitions of Eqs. (6)-(9),   
\begin{multline}
a = ( p_{a}, q_{a}, \gamma_{a}, \nu_{a}, \alpha_{a}, \beta_{a}, m_{a}) , \quad 
b = ( p_{b}, q_{b}, \gamma_{b}, \nu_{b}, \alpha_{b}, \beta_{b}, m_{b} ) , \\
K_{\mu, \mu^{ \prime}, a }^{\sigma} (z) = \int_{ 1}^{ z} \int_{ - 1}^{ 1} d \xi d \eta        (\xi^{2} - \eta^{2})    \xi^{p_a}    \eta^{q_a}   (\xi^{2} - 1 )^{\gamma_{a} / 2} \\
 \times (1 - \eta^{2})^{\nu_{a} / 2} e^{ - \alpha_{a} \xi} e^{+ \beta_{a} \eta}
P_{\mu}^{\sigma} (\xi) P_{\mu^{ \prime}}^{\sigma}(\eta)  
\end{multline}
\begin{multline}
F_{\mu}^{\sigma}(z) = \frac{ - d}{dz} \left( \frac{Q_{l}^{m}(z)}{P_{l}^{m}(z)}   \right)   = \frac{( - 1)^{\sigma}(\mu + \sigma)! / (\mu - \sigma)! }{[ P_{\mu}^{\sigma}(z) ]^{2} (z^{2} - 1) }
\end{multline}
\begin{multline}
Z_{\mu}^{\sigma} = ( - )^{\sigma} \left[ \frac{(\mu - | \sigma | ) ! }{(\mu + | \sigma | ) ! }   \right]^{2} 
\end{multline}
\begin{multline}
\langle 1 / r_{12}  \rangle = \langle  \Phi_{a}(1) ( 1 / r_{12} ) \Phi_{b}(2)    \rangle = \frac{1}{8} \pi^{2} R^{5} \delta(m_{a} + m_{b} ; 0) \\
\times \sum_{\mu = \sigma = | m_{a} |} ^{ \infty} ( 2\mu + 1 ) Z_{\mu}^{\sigma} 
\int_{ 1}^{\infty}  F_{\mu}^{\sigma}(z)  K_{\mu,  \mu,  a}^{\sigma} (z)
K_{\mu,  \mu^,  b}^{\sigma} (z) dz
\end{multline}
For $\frac{1}{2} (\gamma_{a} + \sigma)$ and $\frac{1}{2} (\nu_{a} + \sigma)$    integers, the one-dimensional integral  $K_{\mu, \mu^{ \prime}, \alpha}^{\sigma} (z)$ can be evaluated analytically for each z and inserted in the numerical integration at the appropriate mesh points.  
\begin{multline} 
K_{\mu, \mu^{ \prime}, a }^{\sigma} (z) =  
\sum_{j = 0}^{ [ \, (\mu - \sigma ) / 2 ] \, } \sum_{r = 0}^{ ( \gamma_{a} + \sigma ) / 2} \sum_{k = 0}^{ [ \, (\mu^{ \prime} - \sigma ) / 2 \, ] } \sum_{t = 0}^{ ( \nu_{a} + \sigma ) / 2 } \\  
\frac{ ( 2 \mu - 2 j ) !  \,  ( 2 \mu^{ \prime } - 2 k ) ! \, }{ 2^ { \mu} j ! \, ( \mu - j ) ! \, ( \mu - \sigma - 2 j ) !\, 2^{ \mu^{ \prime}} k !  } \\ \times  \frac{   ( - 1)^{ [ ( \nu_{a} + \sigma ) / 2 ]  - r } [ \frac{1}{2} (\gamma_{a} + \sigma ) ] !  \, ( - 1 )^{t}  [ \frac{1}{2} (\nu_{a} + \sigma ) ] ! \,   }{  ( \mu^{ \prime } - k ) ! \, ( \mu^{ \prime} - \sigma - 2 k ) ! \,  r ! \,  [ \frac{1}{2} (\gamma_{a} + \sigma ) - r ] ! \, t ! \, [ \frac{1}{2} (\nu_{a} + \sigma ) - t ] ! } 
 \\  \times \lbrace   S_{2} ! \, V_{0} ! \, \sum_{s = 0}^{S_{2}} \sum_{\nu = 0}^{V_0} \frac{ [e^{-\alpha_{a}} -  e^{-\alpha_{a} z } z^{S_{2} - s } ] }{\alpha_{a}^{ s + 1} (S_{2} - s) ! \,  }   \frac{ [ e^{\beta_{a} } (- 1 )^{\nu} - e^{-\beta_{a}} (- 1 )^{V_0}   ] }{ \beta_{a}^{ \nu + 1} (V_{0} - \nu ) !   } \\
 -  S_{0} ! \, V_{2} ! \, \sum_{s = 0}^{S_{0}} \sum_{\nu = 0}^{V_2} \frac{ [e^{-\alpha_{a}} -  e^{-\alpha_{a} z } z^{S_{0} - s } ] }{\alpha_{a}^{ s + 1} (S_{0} - s) ! \,  }   \frac{ [ e^{\beta_{a} } (- 1 )^{\nu} - e^{-\beta_{a}} (- 1 )^{V_2}   ] }{ \beta_{a}^{ \nu + 1} (V_{2} - \nu ) !   }  \rbrace , \\
S_{0} = \mu - \sigma + p_{a} - 2 j + 2 r , \quad  S_{2} = S_{0} + 2 , \\
V_{0} = \mu^{\prime} - \sigma + q_{a} - 2 k + 2 t , \quad  V_{2} = V_{0} + 2    \end{multline} 
The upper limit of $j$ is $\frac{1}{2} ( \mu - \sigma )$ or  $\frac{1}{2} ( \mu - \sigma - 1 )$ , whichever is an integral.  The upper limit of $k$ is            $\frac{1}{2} ( \mu^{\prime} - \sigma )$  or  $\frac{1}{2} ( \mu^{\prime} - \sigma - 1  )$ , whichever is an integral.  The upper limit of $t$ is  $\frac{1}{2} ( \nu_{a} + \sigma )$ and the upper limit of r is   $\frac{1}{2} ( \gamma_{a} +  \sigma )$ ; if these are not integrals, the summations are infinite ones.  In practice, the $ K_{\mu, \mu^{ \prime}, \alpha}^{\sigma} (z)$  are evaluated recursively and $\text{numerically}^{ 24, 25} $.  If the $\Phi_{s}(j)$ of Eq. (10) are Slater-type orbitals, the integrals can be $\text{ reexpressed}^{28} $ as a sum of "charge distributions".  The $r_{1 2}$ expansion is needed for the evaluation of $r_{1 2}$ and for raising the value of n in $r_{1 2}^{n}$.  This expansion  [ Eq. (12) ] has been derived by  $\text{Harris}^{27}$.  The partial integration of Eq. (14) is used in the evaluation of the corresponding integral [Eq. (15)].  In Eq. (14), for $z \rightarrow  \infty$ and $\mu \neq 0 $, the first term approaches $0$.  For $\mu = 0$ , $X_{ \mu}^{ \sigma} = 0 $;  $O_{per} \binom{a }{b} $ denotes the interchange of $a$ and $b$ charge distributions.  We have     
\begin{multline} 
 r_{1 2} = \frac{R}{2} \sum_{ \mu = 0 }^{ \infty} \sum_{ \sigma = - \mu }^{ \mu}   \lbrace  (  U_{ \mu}^{ \sigma} g_{ \mu}^{ \sigma} 
+ V_{ \mu}^{ \sigma} h_{ \mu}^{ \sigma }
 + 2 W_{ \mu }^{ \sigma } ) Q_{ \mu }^{ | \sigma | } ( \xi_{ 2  > 1 } )  \\ 
+  \frac{ X_{ \mu }^{ \sigma } \xi_{ 2  > 1 } }{ P_{ \mu }^{ | \sigma | }( \xi_{ 2  > 1 } ) }   \rbrace P_{\mu}^{ | \sigma | }( \xi_{1 < 2} ) P_{\mu}^{ | \sigma | }( \eta_{1} )P_{\mu}^{ | \sigma | }( \eta_{2}  ) e^{i \sigma ( \phi_{1} - \phi_{2} ) } , \\ 
 g_{ \mu}^{ \sigma } = \frac{P_{ \mu + 2 }^{ | \sigma | }( \xi_{1} ) }{P_{ \mu  }^{ | \sigma | }( \xi_{1} ) } + \frac{P_{ \mu + 2 }^{ | \sigma | }( \xi_{2} ) }{P_{ \mu  }^{ | \sigma | }( \xi_{2} ) } + \frac{P_{ \mu + 2 }^{ | \sigma | }( \eta_{1} ) }{P_{ \mu  }^{ | \sigma | }( \eta_{1} ) } + \frac{P_{ \mu + 2 }^{ | \sigma | }( \eta_{2} ) }{P_{ \mu  }^{ | \sigma | }( \eta_{2} ) } , \\ 
h_{ \mu}^{ \sigma } = \frac{P_{ \mu - 2 }^{ | \sigma | }( \xi_{1} ) }{P_{ \mu  }^{ | \sigma | }( \xi_{1} ) } + \frac{P_{ \mu - 2 }^{ | \sigma | }( \xi_{2} ) }{P_{ \mu  }^{ | \sigma | }( \xi_{2} ) } + \frac{P_{ \mu - 2 }^{ | \sigma | }( \eta_{1} ) }{P_{ \mu  }^{ | \sigma | }( \eta_{1} ) } + \frac{P_{ \mu - 2 }^{ | \sigma | }( \eta_{2} ) }{P_{ \mu  }^{ | \sigma | }( \eta_{2} ) }
\end{multline}  
\begin{multline}
U_{ \mu }^{ \sigma } = Z_{ \mu }^{ \sigma } \frac{ ( \mu - | \sigma |  + 1 ) ( \mu - | \sigma | + 2 ) }{  ( 2 \mu + 3 )^{2 }  } , \\
V_{ \mu}^{ \sigma } = - Z_{ \mu }^{ \sigma } \frac{ ( \mu + | \sigma | - 1 ) ( \mu + | \sigma | ) }{ ( 2 \mu - 1 )^{2} } , \\
W_{ \mu }^{ \sigma } = Z_{ \mu }^{ \sigma } \frac{ 2 (2 \mu + 1 ) ( 4 \sigma^{2} - 1 ) }{ ( 2 \mu - 1 )^{2} ( 2 \mu + 3 )^{2} } , \\
X_{ \mu }^{ \sigma } = - \frac{ ( \mu - | \sigma | ) ! \, 2 ( 2 \mu + 1 ) }{ ( \mu + | \sigma | ) ! \, ( 2 \mu - 1 ) ( 2 \mu + 3 ) } ,   
\\  G_{ \mu }^{ \sigma } = \frac{ - d }{ d z } \lbrace \frac{ z }{ [ P_{ \mu }^{ \sigma }( z ) ]^{2} }    \rbrace  = \frac{ - l_{ \mu }^{ \sigma } ( z )  }  { [ P_{ \mu }^{ \sigma }( z ) ]^{2} ( z^{2} - 1 ) } , \\ l_{ \mu }^{ \sigma } ( z ) = ( 2 \mu + 3 ) z^{2} - 2 ( \mu - \sigma + 1 ) z \frac{ P_{ \mu + 1 }^{ \sigma }( z ) }{  P_{ \mu }^{ \sigma }( z ) } \,  - 1 ,
\end{multline}
\begin{multline}
X_{ \mu }^{ \sigma} \int_{ 1 }^{ z } \frac{ x }{ P_{ \mu }^{ \sigma }( x ) } w( x ) d x =  \frac{X_{ \mu }^{ \sigma } z }{ [ P_{ \mu }^{ \sigma }( z ) ]^{ 2 } } \int_{ 1 }^{ z }  P_{ \mu }^{ \sigma }( x ) w( x ) d x  \\ 
 - X_{ \mu }^{ \sigma } \int_{ 1 }^{ z } \frac{ l_{ \mu }^{ \sigma } ( x ) d x }{ \left[ P_{ \mu }^{ \sigma }( x ) \right]^{ 2 } ( x^{ 2 } - 1 ) } \int_{ 1 }^{ x } P_{ \mu }^{ \sigma  }( \xi ) w( \xi ) d \xi   
\end{multline}
\begin{multline}
\langle  r_{12}  \rangle = \langle  \Phi_{a}(1)  \Phi_{b}(2) r_{12}    \rangle = \frac{1}{32} \pi^{2} R^{7} \delta(m_{a} + m_{b} ; 0)  \\  \times  \left[ 1 + O_{per} \binom{a}{b} \right] \sum_{\mu = \sigma = | m_{a} |} ^{ \infty}  \int_{ 1}^{\infty}  d z  K_{\mu,  \mu,  a}^{\sigma} (z) \, [ \,  F_{\mu}^{\sigma}(z)  \widetilde{ K }_{\mu ,  \mu ,  b}^{\sigma} (z)  \\ 
 + \frac{1}{2} X_{ \mu }^{ \sigma } G_{ \mu }^{ \sigma }( z )  K _{\mu ,  \mu ,  b }^{ \sigma } (z) \, ]
\end{multline}
\begin{multline}
\widetilde{ K }_{\mu ,  \mu ,  a}^{\sigma} (z) = U_{ \mu }^{ \sigma } \, [ \,  K_{\mu + 2 ,  \mu ,  a}^{\sigma} (z) +  K_{\mu ,  \mu + 2 ,  a}^{\sigma} (z) \, ]\,   \\  + V_{ \mu }^{ \sigma } \, [ \,  K_{\mu - 2 ,  \mu ,  a}^{\sigma} (z) +  K_{\mu ,  \mu - 2  ,  a}^{\sigma} (z) \, ] \, + W_{ \mu }^{ \sigma } K_{\mu ,  \mu ,  a}^{\sigma} (z)  
\end{multline}
\begin{center} THREE-ELECTRON INTEGRALS \end{center}
The three - and four - electron integrals have been formulated in a straightforward manner, using partial integration.  Equation (17) illustrates a technique of partial integration useful in the derivations.  Whenever the product of two or more associated Lagendre polynomials occurs, these can be replaced by a sum over a single associated Legendre polynomial.  The coefficients involve products of Clebsch - Gordon coefficients.  Equation (22) is a Clebsch - Gordon series.  For further information see Refs. 23 and 28 - 35.  We have : 
\begin{multline}
\int_{ 1 }^{ \infty } f( y ) \, d y  \int_{ t = y }^{ \infty } g( t ) \, d t = \int_{ 1 }^{ \infty } g( y ) \, d y \int_{ t = 1 }^{ y } f( t ) \,  d t ,
\\ u( y ) = \int_{ 1 }^{ y } f( t ) \, d t  ,  \quad  v( y ) = \int_{ 1 }^{ y } g( s ) \, d s  
\end{multline} 
\begin{multline}
\langle  r_{12} r_{13}  \rangle = \langle  \Phi_{a}(1)  \Phi_{b}(2) \Phi_{c}(3)  r_{12} r_{13}    \rangle  =   \\  \frac{1}{128} \pi^{3} R^{11} \delta(m_{a} + m_{b} + m_{c} ; 0) 
 \left[ 1 + O_{per} \binom{b}{c} \right] \sum_{ \mu = \sigma = | m_{b} |}^{ \infty}   \sum_{ \mu^{ \prime } = \sigma^{ \prime } = | m_{c} |}^{ \infty} 
\\ \times \lbrace   \int_{ 1}^{ \infty}  d z \, [ \widetilde{ \mathfrak{ K }}^{ \sigma^{ \prime}}_{ \mu^{ \prime}, \mu^{ \prime}, c }(z)  +  X_{ \mu^{ \prime }}^{ \sigma^{ \prime} } \bar{ \mathfrak{ K }}^{ \sigma^{ \prime}}_{ \mu^{ \prime}, \mu^{ \prime}, c }(z) ]   
\\  \times [ F_{\mu}^{\sigma}(z)  \widetilde{ N }_{ ( \mu,  \mu^{ \prime } ) ,  a }^{ ( \sigma , \sigma^{ \prime } )  } (z) 
\,  K_{\mu ,  \mu ,  b}^{ \sigma } (z) +  F_{\mu}^{\sigma}(z)  N_{  \mu,  \mu^{ \prime }  ,  a }^{  \sigma , \sigma^{ \prime }   } (z) 
\,   \widetilde{ K }_{\mu ,  \mu ,  b}^{ \sigma } (z)  
  \\  +  X_{ \mu }^{ \sigma } G_{ \mu }^{ \sigma }( z ) 
 N_{  \mu,  \mu^{ \prime }  ,  a }^{  \sigma , \sigma^{ \prime }   } (z) 
\,  K_{\mu ,  \mu ,  b}^{ \sigma } (z) ] \,  \\  +  \int_{ 1}^{ \infty}  d z \,   \mathfrak{ K }^{ \sigma^{ \prime}}_{ \mu^{ \prime}, \mu^{ \prime}, c }(z)       [ F_{\mu}^{\sigma}(z)  \overset{ \approx }{ N }_{  \mu,  \mu^{ \prime }  ,  a }^{  \sigma , \sigma^{ \prime }   } (z) 
\,  K_{\mu ,  \mu ,  b}^{ \sigma } (z)   \\  +  F_{\mu}^{\sigma}(z) \widetilde{  N }_{ ( \mu^{ \prime } ,  \mu ) ,  a }^{ (  \sigma^{ \prime } , \sigma )  } (z) \,   \widetilde{ K }_{\mu ,  \mu ,  b}^{ \sigma } (z)  
  +  X_{ \mu }^{ \sigma } G_{ \mu }^{ \sigma }( z )
 \widetilde{  N }_{ ( \mu^{ \prime } ,  \mu ) ,  a }^{ (  \sigma^{ \prime } , \sigma )  } (z) \,  K_{\mu ,  \mu ,  b}^{ \sigma } (z) ] \,   \rbrace 
\end{multline}

\begin{multline}
 \mathfrak{ K }^{ \sigma }_{ \mu, \mu^, b }(z) = \int_{ z }^{ \infty }  F_{\mu}^{\sigma}(z)  \,  K_{\mu ,  \mu ,  b}^{ \sigma } (z) \, d z      
\end{multline}

\begin{multline}
 \bar{\mathfrak{  K } }^{ \sigma }_{ \mu, \mu^, b }(z) = \int_{ z }^{ \infty }  G_{\mu}^{\sigma}(z)  \,  K_{\mu ,  \mu ,  b}^{ \sigma } (z) \, d z      
\end{multline}

\begin{multline}
\widetilde{ \mathfrak{ K }}^{ \sigma }_{ \mu, \mu^, b }(z) = \int_{ z }^{ \infty }  F_{\mu}^{\sigma}(z)  \,\widetilde{ K }_{\mu ,  \mu ,  b}^{ \sigma } (z) \, d z      
\end{multline}
 
\begin{multline}
P_{ l }^{ m } (z) P_{ l^{ \prime } }^{ m^{ \prime } }(z) = \sum_{ j } \left[ 
\begin{matrix} l & l^{ \prime } & j \\
m & m^{ \prime } & ( m + m^{ \prime } ) 
\end{matrix} 
\right] P_{ j }^{ ( m + m^{ \prime } ) } (z)   
\end{multline}

\begin{multline}
\left[ 
\begin{matrix} j_{ 1 } & j_{ 2 } & j \\ m_{ 1 } & m_{ 2 } & m  \end{matrix}       \right]
   =   \left[ 
\begin{matrix} j_{ 2 } & j_{ 1 } & j \\ m_{ 2 } & m_{ 1 } & m  \end{matrix}       \right] = \delta ( m ; m_{ 1 } + m_{ 2 } )  
\\ \times \left[ \frac{ ( j - m ) ! \, (j_{1} + m_{1} ) ! \, (j_{2} + m_{2}) !  \, }{ ( j + m ) ! \, (j_{1} - m_{1} ) ! \, (j_{2} -  m_{2}) !  \, } \right]^{ 1 / 2 } C( j_{1}, j_{2}, j ; m_{1}, m_{2}, m ) \\ \times  C( j_{1}, j_{2}, j ; 0, 0, 0 ) 
\\ = \delta (m ; m_{1} + m_{2} ) ( 2 j + 1 )^{1/2} \Delta (j_{1}, j_{2}, j) \\  \times     \sum_{ p } \frac{ ( -1 )^{p}}{p ! ( j_{1} + j_{2} -j - p ) ! ( j_{1} - m_{1} - p ) ! }  \\
\frac{  [ (j + m) ! ( j - m ) ! (j_{1} + m_{1}) ! (j_{1} - m_{1}) ! (j_{2} + m_{2}) ! (j_{2} - m_{2}) ! ]^{1/2} }{ (j - j_{2} + m_{1} + p ) ! (j_{2} + m_{2} - p ) ! ( j - j_{1} - m_{2} + p ) ! }  ,   
\\ \Delta ( a, b, c ) =  \left[ \frac{ ( a + b - c ) ! ( b + c - a ) ! ( c + a -b ) ! }{ ( a + b + c + 1 ) ! } \right]^{ 1 / 2 }  ,
\\  N^{ \sigma, \sigma^{ \prime } }_{ \mu , \mu^{ \prime }, a } ( z ) 
 =   N^{ \sigma^{ \prime } , \sigma }_{ \mu^{ \prime } , \mu , a } ( z ) = \int_{1}^{ z } \int_{ - 1 }^{ 1 } ( \xi^{2} - \eta^{2} ) d \xi d \eta  
 P_{\mu}^{\sigma} (\xi)  P_{\mu^{ \prime } }^{\sigma^{ \prime } } (\xi)          \\ \times P_{ \mu }^ {\sigma}(\eta) P_{\mu^{ \prime}}^{ \sigma^{ \prime } } (\eta)  \xi^{p_a}    \eta^{q_a}   (\xi^{2} - 1 )^{\gamma_{a} / 2} 
  (1 - \eta^{2})^{\nu_{a} / 2} e^{ - \alpha_{a} \xi} e^{+ \beta_{a} \eta}
\\ = \frac{1}{2} \delta ( m ; \sigma + \sigma^{ \prime } ) \sum_{ J } \sum_{ J^{ \prime } }   \left[ \begin{matrix}  \mu & \mu^{ \prime } & J \\ \sigma & \sigma^{ \prime } & m  \end{matrix} \right]  \left[ \begin{matrix}  \mu & \mu^{ \prime } & J^{ \prime } \\ \sigma & \sigma^{ \prime } & m  \end{matrix} \right] 
\\ \times \left\{ K^{ m }_{ J. J^{ \prime } , a } (z ) + K^{ m }_{ J^{ \prime }, J, a } ( z )      \right\}
\end{multline}
\begin{multline}
\widetilde{ N }^{ ( \sigma, \sigma^{ \prime } ) }_{ ( \mu, \mu^{\prime} ) , a } ( z ) = \delta ( m ; \sigma + \sigma^{ \prime } ) \sum_{ J } \sum_{ J^{ \prime } } \, \lbrace \,  U_{ \mu }^{ \sigma } \left[ \begin{matrix} \mu + 2 & \mu^{ \prime } & J \\ \sigma & \sigma^{ \prime } & m \end{matrix} \right] \\  +   V_{ \mu }^{ \sigma } \left[ \begin{matrix} \mu - 2 & \mu^{ \prime } & J \\ \sigma & \sigma^{ \prime } & m \end{matrix} \right]  \rbrace  \left[ \begin{matrix} \mu  & \mu^{ \prime } & J^{ \prime } \\ \sigma & \sigma^{ \prime } & m \end{matrix} \right]  \lbrace K^{ m }_{ J, J^{ \prime } , a } (z) + K^{ m }_{ J^{ \prime } , J, a } (z) \rbrace \\   +   W^{ \sigma }_{ \mu } N^{ \sigma, \sigma^{ \prime }}_{ \mu, \mu^{ \prime }, a } (z)     
\end{multline} 
\begin{multline}
\overset{ \approx }{ N }^{  \sigma, \sigma^{ \prime }  }_{  \mu, \mu^{\prime}  , a } ( z ) =  \overset{ \approx }{ N }^{  \sigma^{ \prime }, \sigma  }_{  \mu,^{ \prime } , \mu  , a } ( z ) =  
\delta ( m ; \sigma + \sigma^{ \prime } ) \sum_{ J } \sum_{ J^{ \prime } } \, \lbrace \, \\  \left(    \right.    U_{ \mu }^{ \sigma }  U_{ \mu^{ \prime } }^{ \sigma^{ \prime }} \left[ \begin{matrix} \mu + 2 & \mu^{ \prime } + 2  & J \\ \sigma & \sigma^{ \prime } & m \end{matrix} \right]    +  V_{ \mu }^{ \sigma }  V_{ \mu^{ \prime } }^{ \sigma^{ \prime }} \left[ \begin{matrix} \mu - 2 & \mu^{ \prime } - 2  & J \\ \sigma & \sigma^{ \prime } & m \end{matrix} \right]  \\  + U_{ \mu }^{ \sigma }  V_{ \mu^{ \prime } }^{ \sigma^{ \prime }} \left[ \begin{matrix} \mu + 2 & \mu^{ \prime } - 2  & J \\ \sigma & \sigma^{ \prime } & m \end{matrix} \right]  +  V_{ \mu }^{ \sigma }  U_{ \mu^{ \prime } }^{ \sigma^{ \prime }} \left[ \begin{matrix} \mu - 2 & \mu^{ \prime } +  2  & J \\ \sigma & \sigma^{ \prime } & m \end{matrix} \right]  \, \left.   \right)    \left[ \begin{matrix} \mu   & \mu^{ \prime } & J^{ \prime }  \\ \sigma & \sigma^{ \prime } & m \end{matrix} \right]  \\  + \left( U_{ \mu }^{ \sigma }  \left[ \begin{matrix} \mu + 2 & \mu^{ \prime } & J \\ \sigma & \sigma^{ \prime } & m \end{matrix} \right]   +  V_{ \mu }^{ \sigma }  \left[ \begin{matrix} \mu - 2 & \mu^{ \prime } & J \\ \sigma & \sigma^{ \prime } & m \end{matrix} \right]  \,  \right) \\ \times \left( U^{ \sigma^{ \prime }}_{ \mu^{ \prime }}      \left[ \begin{matrix} \mu  & \mu^{ \prime } + 2  & J \\ \sigma & \sigma^{ \prime } & m \end{matrix} \right]  + V^{ \sigma^{ \prime }}_{ \mu^{ \prime }}      \left[ \begin{matrix} \mu  & \mu^{ \prime } - 2  & J \\ \sigma & \sigma^{ \prime } & m \end{matrix} \right]  \right)  \,  \rbrace   \\ \times \lbrace K^{ m}_{ J, J^{ \prime }, a } ( z ) + K^{ m }_ { J^{ \prime } , J, a } ( z ) \rbrace    + W_{ \mu^{ \prime } }^{ \sigma^{ \prime } } \widetilde{ N }^{ ( \sigma, \sigma^{ \prime } ) }_{ ( \mu , \mu^{ \prime } ) , a } ( z ) \\ +  W_{ \mu}^{ \sigma } \widetilde{ N }^{ ( \sigma^{ \prime } , \sigma ) }_{ ( \mu^{ \prime } , \mu ), a } ( z )     -  W_{ \mu}^{ \sigma } W_{ \mu^{ \prime } }^{ \sigma^{ \prime } }  N^{  \sigma, \sigma^{ \prime }  }_{  \mu , \mu^{ \prime }  , a } ( z )                                                      \end{multline}
\begin{multline}
\langle   r_{ 1 2 } /  r_{ 1 3 }   \rangle  = \langle \Phi_{ a } ( 1 ) \Phi_{ b } ( 2 ) \Phi_{ c } ( 3 )  r_{ 1 2 }  /  r_{ 1 3 }  \rangle  =   
 \frac{1}{ 64 } R^{ 9 }  \pi^{ 3 }   \\ \times  \delta ( m_{ a } + m_{ b } + m_ { c } ; 0 )  \sum_{ \mu = \sigma = | m_{b} | }^{ \infty } \sum_{ \mu^{ \prime } = \sigma^{ \prime } = | m_{c} | }^{ \infty } ( 2 \mu^{ \prime } + 1 ) Z^{ \sigma^{ \prime } }_{ \mu^{ \prime } } \lbrace \int_{ 1 }^{ \infty } d z \, F^{ \sigma}_{ \mu }( z ) \\ \times  \mathfrak{ K }^{ \sigma^{ \prime } }_{ \mu^{ \prime }, \mu^{ \prime }, c } ( z ) \left[ \, \widetilde{ N }^{ ( \sigma, \sigma^{ \prime } ) }_{ ( \mu , \mu^{ \prime } ), a } ( z ) K^{ \sigma }_{ \mu. \mu , b } ( z )  +  N ^{  \sigma, \sigma^{ \prime }  }_{  \mu , \mu^{ \prime } , a } ( z ) \widetilde{ K }^{ \sigma }_{ \mu. \mu , b } ( z ) \, \right]  \\  + \int_{ 1 }^{ \infty } d z \, F^{ \sigma^{ \prime } }_{ \mu^{ \prime } } ( z ) K^{ \sigma^{ \prime } }_{ \mu^{ \prime }, \mu^{ \prime } , c } ( z )   \left[ \, \widetilde{ N }^{ ( \sigma, \sigma^{ \prime } ) }_{ ( \mu , \mu^{ \prime } ), a } ( z ) \mathfrak{ K }^{ \sigma }_{ \mu. \mu , b } ( z )  +  N ^{  \sigma, \sigma^{ \prime }  }_{  \mu , \mu^{ \prime } , a } ( z ) \widetilde{ \mathfrak{K} }^{ \sigma }_{ \mu. \mu , b } ( z ) \, \right] 
\\  + X^{ \sigma }_{ \mu } \int_{ 1 }^{ \infty } d z \, N^{ \sigma, \sigma^{ \prime } }_{ \mu , \mu^{ \prime } , a } ( z ) \, \left[ \, K^{ \sigma }_{ \mu , \mu, b } ( z ) \mathfrak{K}^{ \sigma^{ \prime } }_{ \mu^{ \prime }, \mu^{ \prime } , c } ( z ) G_{ \mu }^{ \sigma } ( z )  \right. \\ \left.  +   \bar{ \mathfrak{K} }^{ \sigma }_{ \mu , \mu, b } ( z )  K^{ \sigma^{ \prime } }_{ \mu^{ \prime }, \mu^{ \prime } , c } ( z ) F_{ \mu^{ \prime } }^{ \sigma^{ \prime } } ( z ) \right] \rbrace
\end{multline}
\begin{multline}
\langle r_{12} r_{13} / r_{23} \rangle = \langle \Phi_{a} (1) \Phi_{b} (2) \Phi_{c} (3) r_{12} r_{13} / r_{23}  \rangle =  \\ \frac{1}{128} R^{10} \pi^{3} \delta (m_{a} + m_{b} + m_{c} ; 0 ) \left[ 1 + O_{per} \binom{b}{c} \right]      \\  \times  \sum^{ \infty }_{ \mu^{ \prime \prime }  =  0 } \sum^{ \mu^{ \prime \prime } }_{ w^{ \prime \prime } = - \mu^{ \prime \prime } }  \sum^{ \infty }_{ \mu^{ \prime } = \sigma^{ \prime } } \sum^{ \infty }_{ \mu = \sigma } \delta ( w^{ \prime } ; m_{c} - w^{ \prime \prime } ) \delta ( | w^{ \prime } | ; \sigma^{ \prime } ) \\ \times \delta ( w ; m_{b} + w^{ \prime \prime } ) \delta ( | w | ; \sigma ) \delta ( | w^{ \prime \prime } | ; \sigma^{ \prime \prime } ) ( 2 \mu^{ \prime \prime } + 1  ) Z^{ \sigma^{ \prime \prime } }_{ \mu^{ \prime \prime } } \\ \times \{\ \int_{ 1 }^{ \infty } d z \mathfrak{ N }^{ \sigma^{ \prime }, \sigma^{ \prime \prime } }_{ \mu^{ \prime }, \mu^{ \prime \prime } , c } ( z ) \left[ N^{ \sigma , \sigma^{ \prime \prime } }_{ \mu . \mu^{ \prime \prime } , b  } (  z ) \widetilde{ N }^{ ( \sigma^{ \prime } , \sigma ) }_{ ( \mu^{ \prime } , \mu ) , a } ( z ) G_{ \mu }^{ \sigma } ( z ) X_{ \mu }^{ \sigma }  \right.  \\ + \widetilde{ N }^{ ( \sigma, \sigma^{ \prime \prime } ) }_{ ( \mu , \mu^{ \prime \prime } ), b  } ( z ) \widetilde { N }^{ ( \sigma^{ \prime } , \sigma   ) }_{ ( \mu^{ \prime } , \mu ) , a } ( z ) F_{ \mu }^{ \sigma } ( z )     \left.   +  N^{  \sigma, \sigma^{ \prime \prime }  }_{ \mu , \mu^{ \prime \prime } , b  } ( z )  \overset{ \approx }{ N }^{ \sigma , \sigma^{ \prime } }_{ \mu , \mu^{ \prime } , a } ( z ) F_{ \mu }^{ \sigma } ( z )  \right]  \\ + \int^{ \infty }_{ 1 } d z \left[ X_{ \mu }^{ \sigma } \bar{ \mathfrak{ N } }^{ ( \sigma , \sigma^{ \prime \prime } ) }_{ ( \mu . \mu^{ \prime \prime } ) , b } ( z )  + \widetilde{ \mathfrak{ N } }^{ ( \sigma , \sigma^{ \prime \prime } ) }_{ ( \mu . \mu^{ \prime \prime } ) , b } ( z ) \, \right] \left[ N^{ \sigma, \sigma^{ \prime } }_{ \mu. \mu^{ \prime } , a } ( z ) N^{ \sigma^{ \prime }, \sigma^{ \prime \prime } }_{ \mu^{ \prime }, \mu^{ \prime \prime } , c } (z )  \right.  \\ \times G_{ \mu^{ \prime } }^{ \sigma^{ \prime} } ( z ) X_{ \mu^{ \prime } }^{ \sigma^{ \prime } } + \widetilde{ N }^{ ( \sigma^{ \prime } , \sigma ) }_{ ( \mu^{ \prime } , \mu ) , a } ( z ) N^{ \sigma^{ \prime } , \sigma^{ \prime \prime } }_{ \mu^{ \prime }, \mu^{ \prime \prime } , c } ( z ) F_{ \mu^{ \prime } }^{ \sigma^{ \prime } } ( z )  \\ + N^{ \sigma , \sigma^{ \prime } }_{ \mu, \mu^{ \prime } , a } ( z ) \widetilde{ N }^{ ( \sigma^{ \prime } , \sigma^{ \prime \prime } ) }_{ ( \mu^{ \prime }, \mu^{ \prime \prime } )  , c } ( z )   F^{ \sigma^{ \prime } }_{ \mu^{ \prime } } ( z )   \, \left.      \right]      \\ + \int_{ 1 }^{ \infty } d z \, F_{ \mu^{ \prime \prime } }^{ \sigma^{ \prime \prime } } ( z ) N^{ \sigma, \sigma^{ \prime \prime } }_{ \mu , \mu^{ \prime \prime } , b } ( z ) N^{ \sigma^{ \prime }. \sigma^{ \prime \prime } }_{ \mu^{ \prime } , \mu^{ \prime \prime } , c } ( z ) \, \left[  \,  \right.  X_{ \mu^{ \prime }}^{ \sigma^{ \prime } }  \overset{\eqsim }{ \mathfrak{N} }^{ ( \sigma , \sigma^{ \prime } ) }_{ ( \mu , \mu^{ \prime } ) , a } ( z )   \\   +  \frac{1}{2}  \overset{\approx }{ \mathfrak{N} }^{ ( \sigma , \sigma^{ \prime } ) }_{ ( \mu , \mu^{ \prime } ) , a } ( z )  +    \frac{1}{2}   X_{ \mu }^{ \sigma } X_{ \mu^{ \prime }}^{ \sigma^{ \prime } }  \widehat{\mathfrak{N} }^{  \sigma , \sigma^{ \prime }  }_{  \mu , \mu^{ \prime }  , a } ( z )      \left.  \right]   \\ + \int_{ 1 }^{ \infty } d z \, F_{ \mu^{ \prime \prime } }^{ \sigma^{ \prime \prime } } ( z )  N^{ \sigma , \sigma^{ \prime \prime } }_{ \mu, \mu^{ \prime \prime } , b } ( z )   \widetilde{ N }^{ ( \sigma^{ \prime }, \sigma^{ \prime \prime } ) }_{ ( \mu^{ \prime }, \mu^{ \prime \prime } ) , c } ( z )  \\  \times   \left[  \widetilde{ \mathfrak{N} }^{ ( \sigma, \sigma^{ \prime }  ) }_{ ( \mu. \mu^{ \prime } ) , a } ( z )  + X_{ \mu }^{ \sigma } \bar{ \mathfrak{N} }^{ ( \sigma, \sigma^{ \prime } ) }_{ ( \mu. \mu^{ \prime } ) , a } ( z )   \right]  \\ + \frac{ 1 }{ 2 } \int_{ 1 }^{ \infty } d z \, F_{ \mu^{ \prime \prime }  }^{ \sigma^{ \prime \prime } } ( z ) \widetilde{ N }^{ ( \sigma, \sigma^{ \prime \prime } ) }_{ ( \mu. \mu^{ \prime \prime } ) , b } ( z ) \widetilde{ N }^{ ( \sigma^{ \prime }, \sigma^{ \prime \prime } ) }_{ ( \mu^{ \prime }, \mu^{ \prime \prime } ) , c } ( z ) \mathfrak{N}^{ \sigma, \sigma^{ \prime } }_{ \mu, \mu^{ \prime } , a } ( z )     \}\   \end{multline}              \begin{multline}  \mathfrak{N}^{ \sigma , \sigma^{ \prime } }_{ \mu, \mu^{ \prime } , a } ( z ) =  \mathfrak{N}^{ \sigma^{ \prime } , \sigma }_{ \mu^{ \prime }, \mu , a } ( z ) = \int_{ z }^{ \infty } F^{ \sigma , \sigma^{ \prime } }_{ \mu , \mu^{ \prime } } ( x ) N^{ \sigma , \sigma^{ \prime } }_{ \mu , \mu^{ \prime } , a } ( x ) \, d x   ,     \end{multline}    \begin{multline} F^{ \sigma, \sigma^{ \prime } }_{ \mu , \mu^{ \prime } } ( x ) = \frac{ - d }{ d x } \left[  \frac{ Q_{ \mu}^{ \sigma } ( x ) }{ P_{ \mu }^{ \sigma } ( x ) }  \frac{ Q_{ \mu^{ \prime } }^{ \sigma^{ \prime } } ( x ) }{ P^{ \sigma^{ \prime } }_{ \mu^{ \prime } } ( x ) }   \right]  , \\  E^{ ( \sigma, \sigma^{ \prime } ) }_{ ( \mu , \mu^{ \prime } ) } ( x ) =  \frac{ - d }{ d x } \left\{  \frac{ x }{ [ P^{ \sigma }_{ \mu } ( x ) ]^{ 2 } } \, \frac{ Q^{ \sigma^{ \prime } }_{ \mu^{ \prime }} ( x ) }{ P^{ \sigma^{ \prime }}_{ \mu^{ \prime }} ( x ) } \right\} ,    \\  G^{  \sigma, \sigma^{ \prime }  }_{  \mu , \mu^{ \prime }  } ( x ) =  \frac{ - d }{ d x } \left\{  \frac{ x }{ [ P^{ \sigma }_{ \mu } ( x ) ]^{ 2 } } \,   \frac{ x }{ [ P^{ \sigma^{ \prime } }_{ \mu^{ \prime } } ( x ) ]^{ 2 } } \,        \right\} ,          \end{multline}      \begin{multline}  \widehat{\mathfrak{N}}^{ \sigma , \sigma^{ \prime } }_{ \mu, \mu^{ \prime } , a } ( z ) = \widehat{\mathfrak{N}}^{ \sigma^{ \prime } , \sigma }_{ \mu^{ \prime }, \mu , a } ( z ) = \int_{ z }^{ \infty } G^{ \sigma , \sigma^{ \prime } }_{ \mu , \mu^{ \prime } } ( x ) N^{ \sigma , \sigma^{ \prime } }_{ \mu , \mu^{ \prime } , a } ( x ) \, d x   ,     \end{multline}    \begin{multline}  \bar{\mathfrak{N}}^{ ( \sigma , \sigma^{ \prime } ) }_{ ( \mu, \mu^{ \prime } ) , a } ( z )  =  \int_{ z }^{ \infty } E^{ ( \sigma , \sigma^{ \prime } ) }_{ ( \mu , \mu^{ \prime } ) } ( x ) N^{ \sigma , \sigma^{ \prime } }_{ \mu , \mu^{ \prime } , a } ( x ) \, d x   ,     \end{multline}    \begin{multline}  \widetilde{ \mathfrak{N }}^{ ( \sigma , \sigma^{ \prime } ) }_{ ( \mu, \mu^{ \prime } ) , a } ( z )  =  \int_{ z }^{ \infty } F^{  \sigma , \sigma^{ \prime }  }_{  \mu , \mu^{ \prime }  } ( x ) \widetilde{ N }^{ ( \sigma , \sigma^{ \prime } ) }_{ ( \mu , \mu^{ \prime } ) , a } ( x ) \, d x   ,     \end{multline}      \begin{multline}  \overset{\simeq}{\mathfrak{ N }}^{ ( \sigma , \sigma^{ \prime } ) }_{ ( \mu, \mu^{ \prime } ) , a } ( z )  =   \overset{\eqsim}{\mathfrak{ N }}^{ ( \sigma , \sigma^{ \prime } ) }_{ ( \mu, \mu^{ \prime } ) , a } ( z )  =  \int_{ z }^{ \infty } E^{ ( \sigma^{ \prime }  , \sigma )  }_{ ( \mu^{ \prime } , \mu )  } ( x ) \widetilde{ N }^{ ( \sigma , \sigma^{ \prime } ) }_{ ( \mu , \mu^{ \prime } ) , a } ( x ) \, d x   ,     \end{multline}       \begin{multline}  \overset{\approx}{\mathfrak{ N }}^{  \sigma , \sigma^{ \prime }  }_{  \mu, \mu^{ \prime }  , a } ( z )  =   \overset{\approx}{\mathfrak{ N }}^{  \sigma^{ \prime } , \sigma }_{  \mu^{ \prime }, \mu , a } ( z )  =  \int_{ z }^{ \infty } F^{  \sigma  , \sigma^{ \prime }   }_{  \mu , \mu^{ \prime } } ( x ) \overset{\approx }{ N }^{  \sigma , \sigma^{ \prime }  }_{  \mu , \mu^{ \prime }  , a } ( x ) \, d x   ,     \end{multline}                                 \begin{center} FOUR - ELECTRON INTEGRALS \end{center}    We have                \begin{multline}   \langle  \frac{ r_{ 1 2 } \, r_{ 1 3 } }{ r_{ 1 4 } } \rangle =   \langle \Phi_{a} ( 1 ) \Phi_{b} ( 2 ) \Phi_{c} ( 3 ) \Phi_{d} ( 4 )  \frac{ r_{ 1 2 } \, r_{ 1 3 } }{ r_{ 1 4 } } \rangle = \\  \frac{ 1 }{ 512 } R^{ 13 } \pi^{ 4 } \delta ( m_{a} + m_{b} + m_{c} + m_{d} ; 0 ) \left[ 1 + O_{ per }  \binom{ b }{ c } \right] \sum^{ \infty }_{ \mu = \sigma = | m_{b} | } \sum^{ \infty }_{ \mu^{ \prime } = \sigma^{ \prime } = | m_{c} | }                           
 \\ \times \sum^{ \infty }_{ \mu^{ \prime \prime } = \sigma^{ \prime \prime } = | m_{d} | }  ( 2 \mu^{ \prime \prime }  + 1 ) Z_{ \mu^{ \prime \prime } }^{ \sigma^{ \prime \prime } }  \left\{ \int_{ 1 }^{ \infty }  d z \overset{\approx }{M}^{ ( \sigma , \sigma^{ \prime } , \sigma^{ \prime \prime } ) }_{ ( \mu . \mu^{ \prime }, \mu^{ \prime \prime } ) , a  } ( z ) \mathfrak{ K }^{ \sigma^{ \prime } }_{ \mu^{ \prime } . \mu^{ \prime } , c } ( z )  \right. \\ \times \left[ F_{ \mu }^{ \sigma } ( z ) K^{ \sigma }_{ \mu , \mu , b } ( z ) \mathfrak{K}^{ \sigma^{ \prime \prime } }_{ \mu^{ \prime \prime } , \mu^{ \prime \prime } , d } ( z ) + \frac{ 1 }{ 2 } F^{ \sigma^{ \prime \prime } }_{ \mu^{ \prime \prime } } ( z ) K^{ \sigma^{ \prime \prime } }_{ \mu^{ \prime \prime } , \mu^{ \prime \prime } , d } ( z ) \mathfrak{K}^{ \sigma }_{ \mu , \mu , b } ( z )  \right]  \\  + \int_{ 1 }^{ \infty }   d z  M^{ \sigma , \sigma^{ \prime } , \sigma^{ \prime \prime } }_{ \mu , \mu^{ \prime } , \mu^{ \prime \prime } , a } ( z ) \mathfrak{K}^{ \sigma^{ \prime \prime } }_{ \mu^{ \prime \prime }, \mu^{ \prime \prime } , d } ( z )   \\ \times \left[ F_{ \mu }^{ \sigma } ( z ) \widetilde{ K }^{ \sigma }_{ \mu , \mu , b } ( z ) + X_{ \mu }^{ \sigma } G_{ \mu }^{ \sigma } ( z ) K^{ \sigma }_{ \mu , \mu , b } ( z )  \right] \left[ X^{ \sigma^{ \prime } }_{ \mu^{ \prime } } \bar{\mathfrak{K}}^{ \sigma^{ \prime } }_{ \mu^{ \prime } , \mu^{ \prime } , c } ( z )  +  \widetilde{\mathfrak{K}}^{ \sigma^{ \prime } }_{ \mu^{ \prime } , \mu^{ \prime } , c } ( z )   \right]    \\   + \frac{ 1 }{ 2 } \int_{ 1 }^{ \infty }  d z  F^{ \sigma^{ \prime \prime } }_{ \mu^{ \prime \prime } } ( z )    M^{ \sigma , \sigma^{ \prime } , \sigma^{ \prime \prime } }_{ \mu , \mu^{ \prime } , \mu^{ \prime \prime } , a } ( z )  K^{ \sigma^{ \prime \prime } }_{ \mu^{ \prime \prime }, \mu^{ \prime \prime } , d } ( z )  \\     \times              \left[ X^{ \sigma }_{ \mu }  \bar{\mathfrak{K}}^{ \sigma }_{ \mu , \mu , b } ( z )  +  \widetilde{\mathfrak{K}}^{ \sigma }_{ \mu , \mu , b } ( z )   \right]                          \left[ X^{ \sigma^{ \prime } }_{ \mu^{ \prime } } \bar{\mathfrak{K}}^{ \sigma^{ \prime } }_{ \mu^{ \prime } , \mu^{ \prime } , c } ( z )  +  \widetilde{\mathfrak{K}}^{ \sigma^{ \prime } }_{ \mu^{ \prime } , \mu^{ \prime } , c } ( z )   \right]  \\ + \int_{ 1 }^{ \infty } d z              \widetilde{ M }^{ ( \sigma^{ \prime } , \sigma , \sigma^{ \prime \prime } ) }_{ ( \mu^{ \prime } . \mu , \mu^{ \prime \prime } ) , a  } ( z ) \mathfrak{ K }^{ \sigma^{ \prime } }_{ \mu^{ \prime } . \mu^{ \prime } , c } ( z )   \mathfrak{K}^{ \sigma^{ \prime \prime } }_{ \mu^{ \prime \prime } , \mu^{ \prime \prime } , d } ( z ) \\ \times   \left[ F_{ \mu }^{ \sigma } ( z ) \widetilde{ K }^{ \sigma }_{ \mu , \mu , b } ( z ) + X_{ \mu }^{ \sigma } G_{ \mu }^{ \sigma } ( z ) K^{ \sigma }_{ \mu , \mu , b } ( z )  \right]   \\ + \int_{ 1 }^{ \infty } d z \widetilde{ M }^{ ( \sigma , \sigma^{ \prime } , \sigma^{ \prime \prime } ) }_{ ( \mu , \mu^{ \prime } , \mu^{ \prime \prime } ) , a } ( z ) \left[ \widetilde{\mathfrak{K} }^{ \sigma^{ \prime } }_{ \mu^{ \prime } , \mu^{ \prime } , c } ( z )   +  X^{ \sigma^ \prime }_{ \mu^{ \prime } }  \bar{\mathfrak{K} }^{ \sigma^{ \prime } }_{ \mu^{ \prime } , \mu^{ \prime } , c } ( z )     \right]     \\ \times \left[ F_{ \mu }^{ \sigma } ( z ) K^{ \sigma }_{ \mu , \mu , b } ( z ) \mathfrak{K}^{ \sigma^{ \prime \prime } }_{ \mu^{ \prime \prime } , \mu^{ \prime \prime } , d } ( z )  +   F^{ \sigma^{ \prime \prime } }_{ \mu^{ \prime \prime } } ( z ) K^{ \sigma^{ \prime \prime } }_{ \mu^{ \prime \prime } , \mu^{ \prime \prime } , d } ( z ) \mathfrak{K}^{ \sigma }_{ \mu , \mu , b } ( z )  \right]   \,  \left.  \right\}                                 \end{multline}                                  

\begin{multline}       M^{ \sigma, \sigma^{ \prime } , \sigma^{ \prime \prime } }_{ \mu , \mu^{ \prime } , \mu^{ \prime \prime } , a } ( z )  =  
\\ = \frac{1}{2} \delta ( m ; \sigma + \sigma^{ \prime } + \sigma^{ \prime \prime }  )   \sum_{ L }  \sum_{ L^{ \prime } }  \sum_{ J } \sum_{ J^{ \prime } }   \left[ \begin{matrix}  \mu & \mu^{ \prime } & J \\ \sigma & \sigma^{ \prime } & ( \sigma + \sigma^{ \prime } )  \end{matrix} \right]  \left[ \begin{matrix}  \mu & \mu^{ \prime } & J^{ \prime } \\ \sigma & \sigma^{ \prime } & ( \sigma + \sigma^{ \prime } )  \end{matrix} \right]   
  \\ \times  \left[ \begin{matrix} J & \mu^{ \prime \prime } & L \\ ( \sigma + \sigma^{ \prime } ) & \sigma^{ \prime \prime } & m \end{matrix} \right]    \left[\begin{matrix} J^{ \prime } & \mu^{ \prime \prime } & L^{ \prime } \\ ( \sigma + \sigma^{ \prime } ) & \sigma^{ \prime \prime } & m \end{matrix} \right]      \left\{ K^{ m }_{ L . L^{ \prime } , a } (z ) + K^{ m }_{ L^{ \prime }, L , a } ( z )      \right\}
\end{multline}
\begin{multline}
\widetilde{ M }^{ ( \sigma, \sigma^{ \prime } , \sigma^{ \prime \prime } ) }_{ ( \mu, \mu^{\prime} . \mu^{ \prime \prime } ) , a } ( z ) =
\widetilde{ M }^{ ( \sigma, \sigma^{ \prime \prime } , \sigma^{ \prime } ) }_{ ( \mu, \mu^{\prime \prime} . \mu^{ \prime } ) , a } ( z ) = \delta ( m ; \sigma + \sigma^{ \prime } + \sigma^{ \prime \prime }  ) \sum_{ L } \sum_{ L^{ \prime } } \sum_{ J } \sum_{ J^{ \prime } } \\    \times      \lbrace \,  U_{ \mu }^{ \sigma } \left[ \begin{matrix} \mu + 2 & \mu^{ \prime } & J \\ \sigma & \sigma^{ \prime } & ( \sigma + \sigma^{ \prime } )  \end{matrix} \right]   +   V_{ \mu }^{ \sigma } \left[ \begin{matrix} \mu - 2 & \mu^{ \prime } & J \\ \sigma & \sigma^{ \prime } &  ( \sigma + \sigma^{ \prime } ) \end{matrix} \right]  \rbrace  \left[ \begin{matrix} \mu  & \mu^{ \prime } & J^{ \prime } \\ \sigma & \sigma^{ \prime } & ( \sigma + \sigma^{ \prime } ) \end{matrix} \right]   \\  \times      \left[ \begin{matrix} J & \mu^{ \prime \prime } & L \\ ( \sigma + \sigma^{ \prime } ) & \sigma^{ \prime \prime } & m \end{matrix} \right]  \, \left[ \begin{matrix} J^{ \prime } & \mu^{ \prime \prime } & L^{ \prime } \\ ( \sigma + \sigma^{ \prime } ) & \sigma^{ \prime \prime } & m \end{matrix} \right]  \\ \times    \lbrace K^{ m }_{ L , L^{ \prime } , a } (z) + K^{ m }_{ L^{ \prime } , L, a } (z) \rbrace    +   W^{ \sigma }_{ \mu } M^{ \sigma, \sigma^{ \prime }, \sigma^{ \prime \prime } }_{ \mu, \mu^{ \prime } \mu^{ \prime \prime }, a } (z)     
\end{multline} 
\begin{multline}
\overset{ \approx }{ M }^{ ( \sigma, \sigma^{ \prime } , \sigma^{ \prime \prime } ) }_{ ( \mu, \mu^{\prime} , \mu^{ \prime \prime } ) a } ( z ) =  \overset{ \approx }{ M }^{ ( \sigma^{ \prime }, \sigma , \sigma^{ \prime \prime } )  }_{ ( \mu,^{ \prime } , \mu , \mu^{ \prime \prime } ) , a } ( z ) =   \delta ( m ; \sigma + \sigma^{ \prime } + \sigma^{ \prime \prime } ) \sum_{ J } \sum_{ J^{ \prime } }  \sum_{ L }  \sum_{ L^{ \prime } } 
 \\  \times  \lbrace \,  \left(      \right.    U_{ \mu }^{ \sigma }  U_{ \mu^{ \prime } }^{ \sigma^{ \prime } } \left[ \begin{matrix} \mu + 2 & \mu^{ \prime } + 2  & J \\ \sigma & \sigma^{ \prime } & ( \sigma + \sigma^{ \prime } ) \end{matrix} \right]    +  V_{ \mu }^{ \sigma }  V_{ \mu^{ \prime } }^{ \sigma^{ \prime } } \left[ \begin{matrix} \mu - 2 & \mu^{ \prime } - 2  & J \\ \sigma & \sigma^{ \prime } & ( \sigma + \sigma^{ \prime } )  \end{matrix} \right]   \\  +    U_{ \mu }^{ \sigma }  V_{ \mu^{ \prime } }^{ \sigma^{ \prime }} \left[ \begin{matrix} \mu + 2 & \mu^{ \prime } - 2  & J \\ \sigma & \sigma^{ \prime } & ( \sigma + \sigma^{ \prime } ) \end{matrix} \right]    +  V_{ \mu }^{ \sigma }  U_{ \mu^{ \prime } }^{ \sigma^{ \prime }} \left[ \begin{matrix} \mu - 2 & \mu^{ \prime } +  2  & J \\ \sigma & \sigma^{ \prime } & ( \sigma + \sigma^{ \prime } )  \end{matrix} \right]  \, \left.   \right)  \\ \times  \left[ \begin{matrix} \mu   & \mu^{ \prime } & J^{ \prime }  \\ \sigma & \sigma^{ \prime } & ( \sigma + \sigma^{ \prime } ) \end{matrix} \right]   + \left( U_{ \mu }^{ \sigma }  \left[ \begin{matrix} \mu + 2 & \mu^{ \prime } & J \\ \sigma & \sigma^{ \prime } & ( \sigma + \sigma^{ \prime } ) \end{matrix} \right]   +  V_{ \mu }^{ \sigma }  \left[ \begin{matrix} \mu - 2 & \mu^{ \prime } & J \\ \sigma & \sigma^{ \prime } & ( \sigma + \sigma^{ \prime } )  \end{matrix} \right]  \,  \right) 
\\ \times \left( U^{ \sigma^{ \prime } }_{ \mu^{ \prime } }        \left[ \begin{matrix} \mu  & \mu^{ \prime } + 2  & J \\ \sigma & \sigma^{ \prime } & ( \sigma + \sigma^{ \prime } ) \end{matrix} \right]  + V^{ \sigma^{ \prime }}_{ \mu^{ \prime }}      \left[ \begin{matrix} \mu  & \mu^{ \prime } - 2  & J \\ \sigma & \sigma^{ \prime } & ( \sigma + \sigma^{ \prime } ) \end{matrix} \right]  \right)  \,  \rbrace  
 \\   \times  \left[ \begin{matrix} J & \mu^{ \prime \prime } & L \\ ( \sigma + \sigma^{ \prime } )  & \sigma^{ \prime \prime } & m \end{matrix} \right]  \left[ \begin{matrix} J^{ \prime } & \mu^{ \prime \prime } & L^{ \prime } \\ ( \sigma + \sigma^{ \prime } ) & \mu^{ \prime \prime } & m \end{matrix}  \right]   \lbrace K^{ m}_{ L, L^{ \prime }, a } ( z ) + K^{ m }_ { L^{ \prime } , L, a } ( z ) \rbrace    
\\   +  W_{ \mu^{ \prime } }^{ \sigma^{ \prime } } \widetilde{ M }^{ ( \sigma, \sigma^{ \prime } , \sigma^{ \prime \prime } ) }_{ ( \mu , \mu^{ \prime } , \mu^{ \prime \prime } ) , a } ( z )  +  W_{ \mu}^{ \sigma } \widetilde{ M }^{ ( \sigma^{ \prime } , \sigma  , \sigma^{ \prime \prime } ) }_{ ( \mu^{ \prime } , \mu , \mu^{ \prime \prime } )  a } ( z )     -  W_{ \mu}^{ \sigma } W_{ \mu^{ \prime } }^{ \sigma^{ \prime } }  M^{  \sigma, \sigma^{ \prime } ,\sigma^{ \prime \prime } }_{  \mu , \mu^{ \prime } , \mu^{ \prime \prime } , a } ( z )    
   \end{multline} 
\begin{multline}
\langle r_{ 2 3 } r_{ 1 4 } / r_{ 1 2 } \rangle = \langle \Phi_{ a } ( 1 ) \Phi_{ b } ( 2 ) \Phi_{ c } ( 3 ) \Phi_{ d } ( 4 )   r_{ 2 3 } r_{ 1 4 } / r_{ 1 2 } \rangle =  \frac{ 1 } { 512 } R^{ 13 } \pi^{ 4 } \delta ( \sigma^{ \prime } ; | m_{ c } | )    \\   \times   \delta ( \sigma^{ \prime \prime } ; | m_{ d } | ) \delta ( \sigma ; | m_{ b } + m_{ c } | ) \delta ( m_{ a } + m_{ b } + m_{ c } + m_{ d } ; 0 ) \sum_{ \mu = \sigma }^{ \infty } \sum_{ \mu^{ \prime } = \sigma^{ \prime } }^{ \infty } \sum^{ \infty }_{ \mu^{ \prime \prime } = \sigma^{ \prime \prime } }  ( 2 \mu + 1 )     \\   \times   Z_{ \mu }^{ \sigma } \int^{ \infty }_{ 1 } d z F_{ \mu }^{ \sigma } ( z ) \left[  \right.  N^{ \sigma^{ \prime } , \sigma }_{ \mu^{ \prime } , \mu , b } ( z )  \bar{ \mathfrak{ K } }^{ \sigma^{ \prime } }_{ \mu^{ \prime } , \mu^{ \prime } , c } ( z ) X_{ \mu^{ \prime }}^{ \sigma^{ \prime } }  + \widetilde{ N }^{ ( \sigma^{ \prime } , \sigma ) }_{ ( \mu^{ \prime } , \mu ) , b } ( z )   \mathfrak{ K }^{ \sigma^{ \prime } }_{ \mu^{ \prime } , \mu^{ \prime } , c } ( z ) \\  +    N^{ \sigma^{ \prime } , \sigma }_{ \mu^{ \prime } , \mu , b } ( z )  \widetilde{ \mathfrak{ K } }^{ \sigma^{ \prime } }_{ \mu^{ \prime } , \mu^{ \prime } , c } ( z )  + X_{ \mu^{ \prime } }^{ \sigma^{ \prime } } \bar{ \mathfrak{ J } }^{ ( \sigma^{ \prime }, \sigma ) }_{ ( \mu^{ \prime } , \mu ) , c , b  } ( z )  +  \widetilde{ \mathfrak{ G } }^{ ( \sigma^{ \prime }, \sigma ) }_{ ( \mu^{ \prime } , \mu ) , c , b  } ( z )  \left. \right]  \\  \times   \left[  \right.  N^{ \sigma^{ \prime \prime  } , \sigma }_{ \mu^{ \prime \prime } , \mu , a } ( z )  \bar{ \mathfrak{ K } }^{ \sigma^{ \prime \prime } }_{ \mu^{ \prime \prime } , \mu^{ \prime \prime } , d } ( z ) X_{ \mu^{ \prime\prime } }^{ \sigma^{ \prime \prime } }  + \widetilde{ N }^{ ( \sigma^{ \prime \prime } , \sigma ) }_{ ( \mu^{ \prime \prime  } , \mu ) , a } ( z )   \mathfrak{ K }^{ \sigma^{ \prime \prime } }_{ \mu^{ \prime \prime } , \mu^{ \prime \prime } , d } ( z ) \\  +    N^{ \sigma^{ \prime \prime } , \sigma }_{ \mu^{ \prime \prime } , \mu , a } ( z )  \widetilde{ \mathfrak{ K } }^{ \sigma^{ \prime \prime  } }_{ \mu^{ \prime \prime } , \mu^{ \prime \prime } , d } ( z )  + X_{ \mu^{ \prime \prime  } }^{ \sigma^{ \prime \prime } } \bar{ \mathfrak{ J } }^{ ( \sigma^{ \prime \prime }, \sigma ) }_{ ( \mu^{ \prime \prime } , \mu ) , d , a  } ( z )  +  \widetilde{ \mathfrak{ G } }^{ ( \sigma^{ \prime \prime }, \sigma ) }_{ ( \mu^{ \prime \prime } , \mu ) , d , a  } ( z )  \left. \right]             \end{multline}   
\begin{multline}    \widetilde{ \mathfrak{ G } }^{ ( \sigma^{ \prime \prime }, \sigma ) }_{ ( \mu^{ \prime \prime } , \mu ) , d , a  } ( z )  =   \int^{ z }_{ 1 } d x F_{ \mu^{ \prime \prime } }^{ \sigma^{ \prime \prime } } ( x ) \left[    K^{ \sigma^{ \prime \prime } }_{ \mu^{ \prime \prime }. \mu^{ \prime \prime } . d } ( x ) \widetilde{ N }^{ ( \sigma^{ \prime \prime } , \sigma ) }_{ ( \mu^{ \prime \prime }, \mu ) , a } ( x )  \right.  \\  \left.  +    \widetilde{ K }^{ \sigma^{ \prime \prime } }_{ \mu^{ \prime \prime }. \mu^{ \prime \prime } . d } ( x )  N^{  \sigma^{ \prime \prime } , \sigma  }_{  \mu^{ \prime \prime }, \mu  , a } ( x )   \right]  \end{multline}
\begin{multline}
\bar{ \mathfrak{ J } }^{ ( \sigma , \sigma^{ \prime } ) }_{ ( \mu. \mu^{ \prime } ) , d, a } ( x ) = \int_{ 1 }^{ z } d x G_{ \mu }^{ \sigma } ( x ) K^{ \sigma }_{ \mu , \mu , d } ( x ) N^{ \sigma , \sigma^{ \prime } }_{ \mu , \mu^{ \prime } , a } ( x )    \end{multline}
\begin{multline}                                   \langle r_{ 2 3 } r_{ 1 2 } / r_{ 1 4 } \rangle = \langle \Phi_{ a } ( 1 ) \Phi_{ b } ( 2 ) \Phi_{ c } ( 3 ) \Phi_{ d } ( 4 )   r_{ 2 3 } r_{ 1 2 } / r_{ 1 4 } \rangle =  \frac{ 1 } { 512 } R^{ 13 } \pi^{ 4 } \delta ( \sigma^{ \prime } ; | m_{ c } | )    \\   \times   \delta ( \sigma^{ \prime \prime } ; | m_{ d } | ) \delta ( \sigma ; | m_{ b } + m_{ c } | ) \delta ( m_{ a } + m_{ b } + m_{ c } + m_{ d } ; 0 ) \sum_{ \mu = \sigma }^{ \infty } \sum_{ \mu^{ \prime } = \sigma^{ \prime } }^{ \infty } \sum^{ \infty }_{ \mu^{ \prime \prime } = \sigma^{ \prime \prime } }  ( 2 \mu^{ \prime \prime } + 1 )     \\   \times   Z_{ \mu^{ \prime \prime }  }^{ \sigma^{ \prime \prime }  }  \left(   \right.   \int^{ \infty }_{ 1 } d z  \left\{  \right.     X_{ \mu }^{ \sigma }  G_{ \mu }^{ \sigma } ( z )  \left[    \mathfrak{ J }^{ ( \sigma^{ \prime \prime } , \sigma ) }_{ ( \mu^{ \prime \prime } , \mu ) ,  d ,  a } ( z )  + N^{ \sigma, \sigma^{ \prime \prime } }_{ \mu , \mu^{ \prime \prime } , a } ( z ) \mathfrak{K}^{ \sigma^{ \prime \prime } }_{ \mu^{ \prime \prime } , \mu^{ \prime \prime } ,  d } ( z ) \right]      \\  +        F_{ \mu }^{ \sigma } ( z )  \left[ \widetilde{ \mathfrak{ J } }^{ ( \sigma , \sigma^{ \prime \prime } ) }_{ ( \mu , \mu^{ \prime \prime }  ) ,  d ,  a } ( z )  +  \widetilde{ N }^{ ( \sigma, \sigma^{ \prime \prime } ) }_{ ( \mu , \mu^{ \prime \prime } ) , a } ( z ) \mathfrak{K}^{ \sigma^{ \prime \prime } }_{ \mu^{ \prime \prime } , \mu^{ \prime \prime } ,  d } ( z ) \right]  \left.  \right\}           \\           \times   \left\{   \right.  N^{ \sigma^{ \prime }, \sigma }_{ \mu^{ \prime } , \mu , b } ( z ) \widetilde{ \mathfrak{ K } }^{ \sigma^{ \prime } }_{ \mu^{ \prime }, \mu^{ \prime } , c } ( z )  +  \widetilde{ N }^{ ( \sigma^{ \prime }, \sigma ) }_{ ( \mu^{ \prime } , \mu ) , b } ( z )    \mathfrak{ K }^{ \sigma^{ \prime } }_{ \mu^{ \prime }, \mu^{ \prime } , c } ( z )   +  X_{ \mu^{ \prime } }^{ \sigma^{ \prime } }    N^{  \sigma , \sigma^{ \prime }  }_{  \mu , \mu^{ \prime }  , b } ( z ) \bar{ \mathfrak{ K } }^{ \sigma^{ \prime } }_{ \mu^{ \prime }, \mu^{ \prime } , c } ( z )     \\  + X_{ \mu^{ \prime } }^{ \sigma^{ \prime } }  \bar{ \mathfrak{ J } }^{ ( \sigma^{ \prime }. \sigma ) }_{ ( \mu^{ \prime } , \mu ) ,c , b  } ( z )  +  \widetilde{ \mathfrak{ G } }^{ ( \sigma^{ \prime } , \sigma ) }_{ ( \mu^{ \prime } , \mu ) , c , b } ( z ) \left. \right\}                     \\  + \int^{ \infty }_ { 1 }  d z         F_{ \mu }^{ \sigma } ( z )  \left[  \mathfrak{J}^{ ( \sigma^{ \prime \prime } , \sigma  ) }_{ (  \mu^{ \prime \prime } , \mu  ) ,  d ,  a } ( z )  +   N^{ \sigma, \sigma^{ \prime \prime }  }_{  \mu , \mu^{ \prime \prime }  , a } ( z ) \mathfrak{K}^{ \sigma^{ \prime \prime } }_{ \mu^{ \prime \prime } , \mu^{ \prime \prime } ,  d } ( z ) \right]            \\  \times     \left[   \right.  \widetilde{ N }^{ ( \sigma , \sigma^{ \prime } ) }_{ ( \mu , \mu^{ \prime } ) , b } ( z ) \widetilde{ \mathfrak{ K } }^{ \sigma^{ \prime } }_{ \mu^{ \prime }, \mu^{ \prime } , c } ( z )  +  \overset{ \approx }{ N }^{ \sigma, \sigma^{ \prime } }_{ \mu , \mu^{ \prime }  , b } ( z )    \mathfrak{ K }^{ \sigma^{ \prime } }_{ \mu^{ \prime }, \mu^{ \prime } , c } ( z )   +  X_{ \mu^{ \prime } }^{ \sigma^{ \prime } }  \widetilde{ N }^{ ( \sigma , \sigma^{ \prime } ) }_{ (  \mu , \mu^{ \prime } )  , b } ( z ) \bar{ \mathfrak{ K } }^{ \sigma^{ \prime } }_{ \mu^{ \prime }, \mu^{ \prime } , c } ( z )     \\  + X_{ \mu^{ \prime } }^{ \sigma^{ \prime } }  \overset{ \eqsim }{ \mathfrak{ J } }^{ ( \sigma , \sigma^{ \prime } ) }_{ ( \mu , \mu^{ \prime }  ) ,c , b  } ( z )  +  \overset{ \approx }{ \mathfrak{ G } }^{ ( \sigma , \sigma^{ \prime }  ) }_{ ( \mu , \mu^{ \prime }  ) , c , b } ( z ) \left. \right] \left. \right)        \end{multline}                                                                 \begin{multline}              \widetilde{ \mathfrak{ J } }^{ ( \sigma^{ \prime \prime } , \sigma ) }_{ ( \mu^{ \prime \prime } , \mu ) , d, a } ( z ) = \int^{ z }_{ 1 } F_{ \mu }^{ \sigma } ( x ) K^{ \sigma }_{ \mu , \mu , d } ( x ) \widetilde{ N }^{ ( \sigma^{ \prime \prime } , \sigma ) }_{ ( \mu^{ \prime \prime }, \mu ) , a } ( x )  d  x  ,                                  \end{multline}         \begin{multline}              \overset{ \eqsim }{ \mathfrak{ J } }^{ ( \sigma^{ \prime \prime } , \sigma ) }_{ ( \mu^{ \prime \prime } , \mu ) , d, a } ( z ) = \int^{ z }_{ 1 } G_{ \mu }^{ \sigma } ( x ) K^{ \sigma }_{ \mu , \mu , d } ( x ) \widetilde{ N }^{ ( \sigma^{ \prime \prime } , \sigma ) }_{ ( \mu^{ \prime \prime }, \mu ) , a } ( x )  d  x  ,                        \end{multline}                                   \begin{multline}              \overset{ \approx }{ \mathfrak{ G } }^{ ( \sigma , \sigma^{ \prime  }  ) }_{ ( \mu , \mu^{  \prime }  ) , c , b } ( z ) = \int^{ z }_{ 1 } F_{ \mu^{ \prime } }^{ \sigma^{ \prime } } ( x ) \left[   K^{ \sigma^{ \prime } }_{ \mu^{ \prime } , \mu^{ \prime } , c } ( x )   \overset{ \approx }{ N }^{  \sigma , \sigma^{ \prime  }   }_{  \mu , \mu^{ \prime }  , b } ( x )  \right.  \\  \left.  +   \widetilde{ K }^{ \sigma^{ \prime } }_{ \mu^{ \prime } , \mu^{ \prime } , c } ( x ) \widetilde{ N }^{ ( \sigma , \sigma^{ \prime  }  ) }_{ ( \mu , \mu^{ \prime } ) , b } ( x )  \right]  d  x  ,                                  \end{multline}                               \begin{multline}              \mathfrak{ J }^{ ( \sigma^{ \prime \prime } , \sigma ) }_{ ( \mu^{ \prime \prime } , \mu ) , d, a } ( z ) = \int^{ z }_{ 1 } F_{ \mu^{ \prime \prime } }^{ \sigma^{ \prime \prime } } ( x ) K^{ \sigma^{ \prime \prime }  }_{ \mu^{ \prime \prime } , \mu^{ \prime \prime } , d } ( x )   N^{  \sigma^{ \prime \prime } , \sigma  }_{  \mu^{ \prime \prime }, \mu  , a } ( x )  d  x  ,                                  \end{multline}                           \begin{center} KINETIC ENERGY INTEGRAL                          \end{center}     Essentially no new basic integrals are involved in the evaluation of the kinetic energy, nuclear attraction, and overlap integrals.  In some cases, a modified $ K^{ \sigma }_{ \mu , \mu^{ \prime } , s } ( z ) $ integral is used.  The modified integral    $ H^{ \sigma }_{ \mu , \mu^{ \prime } , s } ( z ) $  is defined in Eq. ( 47); it differs in that the $ \xi^{ 2 } - \eta^{ 2 } $ term is not included.    \begin{multline}       H_{ \mu. \mu^{ \prime } , s }^{ \sigma } ( z ) =  \int_{ 1 }^{ z } \int_{ -1 }^{ 1 } \xi^{ p_{ s } }  \eta^{ q_{ s } } ( \xi^{ 2 } - 1 )^{ \gamma_{ s } / 2 } ( 1 - \eta^{ 2 } )^{ \nu_{ s } /2 }  \\  \times     e^{ - \alpha_{ s } \xi  } e^{ \beta_{ s } \eta } P_{ \mu }^{ \sigma } ( \xi ) P_{ \mu^{ \prime } }^{ \sigma } ( \eta )   d \xi  d  \eta   ,           \end{multline}       \begin{multline}    -   \frac{ 1 }{ 2 } \int  d \tau  \Phi_{ s } ( 1 ) \Phi_{ t } ( 2 ) \Phi_{ w } ( 3 ) r_{ 1 3 }^{ l^{ \prime } }  \nabla^{ 2 }_{ 1 } \left[   r^{ l }_{ 1 2 } \Phi_{ a } ( 1 ) \Phi_{ x } ( 2 )  \Phi_{ y } ( 3 ) \right]  =  \\  - \frac{ 1 }{ 2 } l ( l + 1 ) \int d \tau  \Phi_{ e } ( 1 ) \Phi_{ f } ( 2 ) \Phi_{ g } ( 3 ) r_{ 1 3 }^{ l^\prime } r_{ 1 2 }^{ l - 2 }            \\     -  \frac{ 2 }{ R^{ 2 } }  \int d \tau \frac{ D_{ a } ( 1 ) \Phi_{ e } ( 1 ) }{ \xi_{ 1 }^{ 2 } - \eta_{ 1 }^{ 2 } } \Phi_{ f } ( 2 ) \Phi_{ g } ( 3 ) r_{ 1 3 }^{ l^{ \prime } } r_{ 1 2 }^{ l }  \\  - l \int d \tau \frac{ V_{ a } ( 1 ,  2 ) \Phi_{ e } ( 1 ) }{ \xi_{ 1 }^{ 2 } - \eta_{ 1 }^{ 2 } }  \Phi_{ f } (2 ) \Phi_{ g } ( 3 ) r_{ 1 3 }^{ l^{ \prime } }  r_{ 1 2 }^{ l - 2 } ,                                      \end{multline}                                                \begin{multline}  \Phi_{ s } ( 1 ) \Phi_{ a } ( 1 ) = \Phi_{ e } ( 1 )  ,       \Phi_{ t } ( 2 ) \Phi_{ x } ( 2 ) = \Phi_{ f } ( 2 )    ,                        \Phi_{ w } ( 3 ) \Phi_{ y } ( 3 ) = \Phi_{ g } ( 3 )  , \\   D_{ a } ( 1 ) =  p_{ a }^{ 2 } + p_{ a } + 2 p_{ a } \gamma_{ a }  + \gamma_{ a } - \alpha_{ a }^{ 2 } - q_{ a }^{ 2 } - q_{ a } -2 q_{ a } \nu_{ a }  - \nu_{ a }  \\   + \beta_{ a }^{ 2 } - 2 \alpha_{ a } \xi_{ 1 } ( p_{ a } + \gamma_{ a } + 1 )  -2 \beta_{ a } \eta_{ 1 } ( q_{ a } + \nu_{ a } + 1 )  + \frac{ p_{ a } - p_{ a }^{ 2 }  }{ \xi_{ 1 }^{ 2 } } + \frac{ q_{ a }^{ 2 } - q_{ a } }{ \eta_{ 1 }^{ 2 } }  \\  + \frac{ 2 \alpha_{ a } p_{ a } }{ \xi_{ 1 } }  + \frac{ 2 \beta_{ a } q_{ a } }{ \eta_{ 1 } } + \alpha_{ a }^{ 2 } \xi_{ 1 }^{ 2 } + \frac{ \gamma_{ a }^{ 2 } \xi_{ 1 }^{ 2 } }{  \xi_{ 1 }^{ 2 } - 1 } - \beta_{ a }^{ 2 } \eta_{ 1 }^{ 2 }  + \frac{ \nu_{ a }^{ 2 } \eta_{ 1 }^{ 2 } }{ 1 - \eta_{ 1 }^{ 2 } }  + \frac{ m_{ a }^{ 2 } ( \eta_{ 1 }^{ 2 } - \xi_{ 1 }^{ 2 } ) }{ ( \xi_{ 1 }^{ 2 } - 1 ) ( 1 - \eta_{ 1 }^{ 2 } ) }   ,  \\  V_{ a } ( 1 , 2 ) =   ( 1 - \eta_{ 1 }^{ 2 } )( q_{ a }  + \beta_{ a } \eta_{ 1 }   - \frac{ q_{ a } \xi_{ 2 } \eta_{ 2 } \xi_{ 1 } }{ \eta_{ 1 } } - \beta_{ a } \xi_{ 2 } \eta_{ 2 } \xi_{ 1 } )     \\  +   ( \xi_{ 1 }^{ 2 } - 1 )( p_{ a } - \alpha_{ a } \xi_{ 1 }  -  \frac{p_{ a } \xi_{ 2 } \eta_{ 2 } \eta_{ 1 } }{ \xi_{ 1 } }  + \alpha_{ a } \xi_{ 2 } \eta_{ 2 } \eta_{ 1 }  )  \\ - \nu_{ a } \eta_{ 1 }^{ 2 } + ( \nu_{ a } - \gamma_{ a } ) \xi_{ 2 } \eta_{ 2 } \xi_{ 1 } \eta_{ 1 }   +  \gamma_{ a } \xi_{ 1 }^{ 2 }  \\             + \left( \alpha_{ a } \xi_{ 1 } + \beta_{ a } \eta_{ 1 }  +  q_{ a } - p_{ a } - \frac{ \eta_{ 1 }^2 \nu_{ a } }{ 1 - \eta_{ 1 }^{ 2 } }  - \frac{ \xi_{ 1 }^{ 2 } \gamma_{ a } }{ \xi_{ 1 }^{ 2 } - 1 }  \right)  \\ \times  \left[ ( \xi_{ 2 }^{ 2 }  -1 ) ( 1 - \eta_{ 2 }^{ 2 } ) ( \xi_{ 1 }^{ 2 } - 1 ) ( 1 - \eta_{ 1 }^{ 2 } ) \right]^{ 1 / 2 }   \\ \times \left[  \frac{ e^{ i ( \phi_{ 1 } - \phi_{ 2 } ) } + e^{ - i ( \phi_{ 1 } - \phi_{ 2 } ) } }{ 2 }   \right]   \\  +  m_{ a }  ( \xi_{ 1 }^{ 2 } - \eta_{ 1 }^{ 2 } ) \left[ \frac{ ( \xi_{ 2 }^{ 2 } - 1 ) ( 1  - \eta_{ 2 }^{ 2 } ) }{ ( \xi_{ 1 }^{ 2 } - 1 ) ( 1 - \eta_{ 1 }^{ 2 } ) }    \right]^{ 1 / 2 } \\  \times  \left[  e^{ i ( \phi_{ 1 } - \phi_{ 2 } )} - e^{ -  i ( \phi_{ 1 } - \phi_{ 2 } ) } \right]   ,                               \end{multline}                                                                 For $ l = l^{ \prime } = 0 $ , the kinetic energy integral [Eq. (48)] over electron 1 is the sum of $ - \frac{1}{2} \pi R H^{ \sigma }_{ \mu , \mu , e } ( \infty )  $ terms.  For $ l = 0 $ and $l^{ \prime } = 1 $ , the integral is a sum of $ \langle r_{ 1 2 } \rangle $ terms [Eq. (15)] ,evaluated using $ H^{ \sigma }_{ \mu , \mu^{ \prime }, e } ( z ) $ [Eq. (47)] for electron 1 and the usual    $ K^{ \sigma }_{ \mu , \mu^{ \prime } , f } ( z ) $ [Eq. (7)] for electron 2.  For $ l = 0 $ and $ l^{ \prime } = 2 $ , use [Eq. (4)] for $ \langle r^{ 2 }_{ 1 2 } \rangle $ with $ H^{ \sigma }_{ \mu, \mu^{ \prime } , e } ( z ) $ for electron 1.  For $ l = 1 $ and $ l^{ \prime } = 0 $ , use the usual $ \langle 1 / r_{ 1 2 } \rangle $ [Eq. (10)] and $ \langle r_{ 1 2 } \rangle $ [Eq. (15)] with [Eq. (47)] instead of [Eq. (7)] for electron 1.  For $ l = l^{ \prime } = 1 $ , use $ \langle r_{ 1 3 } / r_{ 1 2 } \rangle $ [Eq. (26)] and $ \langle r_{ 1 2 } r_{ 1 3 } \rangle $  [Eq. (18)] with [Eq. (47)] instead of [Eq. (7)] for electron 1.  For $ l = 2 $ and $ l^{ \prime } = 0 $ , [Eq. (49)] is the sum of         \begin{equation*} - \frac{ 3 }{ 4 }  \pi  R^{ 3 } ( \frac{ 1 }{ 4 } \pi R^{ 3 } )^{ 2 } K^{ \sigma }_{ \mu , \mu , e } ( \infty ) K_{ \mu^{ \prime } , \mu^{ \prime } , f }^{ \sigma^{ \prime } } ( \infty ) K^{ \sigma^{ \prime \prime } }_{ \mu^{ \prime \prime }, \mu^{ \prime \prime } , g } ( \infty ) \end{equation*}     terms, \begin{equation*}  -  \frac{ 1 }{ 2 } \pi R K^{ \sigma }_{ \mu , \mu , g } ( \infty ) \langle r^{ 2 }_{ 1 2 } \rangle       \end{equation*}    terms, and  \begin{equation*} - \frac{ 1 }{ 2 }  \pi  R^{ 3 } ( \frac{ 1 }{ 4 } \pi R^{ 3 } )^{ 2 } H^{ \sigma }_{ \mu , \mu , e } ( \infty ) K_{ \mu^{ \prime } , \mu^{ \prime } , f }^{ \sigma^{ \prime } } ( \infty ) K^{ \sigma^{ \prime \prime } }_{ \mu^{ \prime \prime }, \mu^{ \prime \prime } , g } ( \infty ) \end{equation*}    terms.  The $ \langle r_{ 1 2 }^{ 2 } \rangle $ are evaluated using [Eq. (47)] instead of [Eq.(7)] for electron 1.  The case of the kinetic energy integral in which the Laplacian operates on $ r_{ 1 2 } $ and this is multiplied by $ r_{ 1 2 } $ is represented in [Eq. (50)].           \begin{multline}                    - \frac{ 1 }{ 2 } \int d \tau \Phi_{ s } ( 1 ) \Phi_{ t } ( 2 ) r_{ 1 2 }^{ l^{ \prime } } \nabla_{ 1 }^{ 2 } [ r_{ 1 2 }^{ l } \Phi_{ a } ( 1 ) \Phi_{ x } ( 2 ) ]  ,    \end{multline}   For $ l = l^{ \prime } = 1 $ [Eq.(50)] equals       \begin{multline*}  ( \frac{ 1 }{ 4 } \pi R^{ 3 } )^{ 2 }  \delta ( m_{ s } + m_{ a } ; 0 ) \delta ( m_{ t } + m_{ x } ; 0 ) K^{ 0 }_{ 0 , 0 , e } ( \infty )      K^{ 0 }_{ 0 , 0 , f } ( \infty )   \\  - \frac{ 1 }{ 2 } \int  d \tau  \frac{ \Phi_{ f } ( 2 ) r_{ 1 2 }^{ 2 } [ D_{ a } ( 1 ) + D_{ s } ( 1 ) ] \Phi_{ g } ( 1 )  }{ ( \frac{ 1 }{ 2 } R  )^{ 2 } ( \xi^{ 2 }_{ 1 } - \eta^{ 2 }_{ 1 } ) }        \end{multline*}                                                            \begin{center}  NUCLEAR - ELECTRON ATTRACTION AND OVERLAP INTEGRALS \end{center}   We have             \begin{multline}  - \frac{ 2 }{ R } \int d \tau \left[   \frac{ ( Z_{ a } + Z_{ b } ) \xi_{ 1 } + ( Z_{ a } - Z_{ b } ) \eta_{ 1 } }{ ( \xi^{ 2 }_{ 1 } - \eta^{ 2 }_{ 1 } ) }  \right]  \Phi_{ e } ( 1 ) \Phi_{ f } ( 2 ) \Phi_{ g } ( 3 ) r_{ 1 2 }^{ l } r_{ 1 3 }^{ l^{ \prime } }  ,                     \end{multline}    For $ l = l^{ \prime } = 0 $ , the integral over electron 1 is  \begin{multline*} - \frac{ 1 }{ 2 } \pi R \delta ( m_{ e } ; 0 ) [ ( Z_{ a } + Z_{ b } ) H^{ 0 }_{ 1 , 0 , e } ( \infty ) + ( Z_{ a } - Z_{ b } ) H^{ 0 }_{ 0 , 1 , e } ( \infty ) ]                         \end{multline*}                 For $l^{ \prime } = 0 $ and $ l = 1 $, use $ \langle r_{ 1 2 } \rangle $  [Eq. (15)]  and [Eq. (47)] instead of [Eq. (7)] for electron 1.  For $l^{ \prime } = 0 $ and $ l = 2 $ , use $ \langle r^{ 2 }_{ 1 2 } \rangle $  [Eq. (4)] without the $ ( \xi_{ 1 }^{ 2 } - \eta_{ 1 }^{ 2 } ) $  term.  For $ l^{ \prime } = l = 1 $  use $ \langle r_{ 1 2 } r_{ 1 3 } \rangle $ [Eq. (18)] with [Eq. (47)] instead of [Eq. (7)] for electron 1.  The nuclear-nuclear repulsion integral is $ Z_{ a } Z_{ b } / R $ times the overlap integral [Eq. (52)] .                      \begin{multline}  \langle r_{ 1 2 }^{ l } r_{ 1 3 }^{ l^{ \prime } } \rangle  = \int d \tau  \Phi_{ e } ( 1 ) \Phi_{ f } ( 2 ) \Phi_{ g } ( 3 ) r_{ 1 2 }^{ l } r_{ 1 3 }^{ l^{ \prime } }  ,       \end{multline}   For $ l = l^{ \prime } = 0 $ the integral over electron 1 is $ \frac{ 1 }{ 4 } \pi R^{ 3 } \delta ( m_{ e } ; 0 ) K^{ 0 }_{ 0 , 0 , e } ( \infty ) $ .  For $ l^{ \prime } = 0 $ and $ l = 1 $ , the integral is $ \langle r_{ 1 2 } \rangle $. [Eq. (15)] . For $ l^{ \prime } = 0 $ and $ l = 2 $ , the integral is $ \langle r_{ 1 2 }^{ 2 } \rangle $ [Eq. (4)] .  For $ l^{ \prime } = l = 1 $ the overlap is  $ \langle r_{ 1 2 } r_{ 1 3 } \rangle $ [Eq. (18)].                                                 \begin{center} ADDITIONAL INTEGRALS \end{center}                                Some integrals can be generated from previously given integrals by raising or lowering the $ r_{ 1 2 } $ index in even or odd steps. [Eqs.  (53) and (54)] , using  Eqns. (1) , (5) , (12), and  (59).  If the Hamiltonian contains the term $ 1 / r_{ i j }^{ 3 } $ , for the evaluation of spin-spin magnetic coupling or the relativistic effects of an external electric field, then $ \langle 1 / r_{ i j }^{ 3 } \rangle $ [Eq. (55)] and $ \langle 1 / r_{ i j }^{ 2 } \rangle $ [Eq. ( 56)] are some of the integrals needed.  These results are based on a generalization of the Neumann expansion $ [Eq. (59)]^{ 23, 31, 36, 37 } $ .  The $ C_{ n }^{ l } (x ) $ [Eq. (58)] are Gegenbauer polynomials.  For $ l = \frac{ 1 }{ 2 } $ , the Gegenbauer polynomials are the same as Legendre polynomials.  If the wave function [Eq. ( 3 )] is modified to be                                            \[
\Psi_{tot} = \tilde{\mathcal{A}} \quad  \left\{ \left[ \prod_j \sum_s a_{s_j} \phi_{s}(j) \right] \left[ 1 + \sum_{j \quad <} \sum_{l \quad <} \sum_{k} w_{jlk} r_{jl}^{n} r_{lk}^{\nu}   \right] \right\} ,
  \]
then  terms $ \langle 1 / r_{ 1 2 } r_{ 1 3 } \rangle $ [Eq. (60)] will occur in the kinetic energy integrals.
\begin{multline} \langle r_{ 1 2 }^{ 3 } \rangle   = \int d \tau \Phi_{ e } ( 1 ) \Phi_{ f } ( 2 ) r_{ 1 2 }^{ 3 } =  \frac{ 1 }{ 4 } R^{ 2 } \left[  \right.   \langle \Phi_{ ( p_{ e } + 2 , \cdots ) } ( 1 ) r_{ 1 2 } \Phi_{ f } ( 2 ) \rangle   \\   + \langle \Phi_{ ( p_{ e } , q_{ e } + 2 , \cdots ) } ( 1 ) r_{ 1 2 } \Phi_{ f } ( 2 ) \rangle   - 2 \langle \Phi_{ e } ( 1 ) r_{ 1 2 } \Phi_{ f } ( 2 ) \rangle  \\  + \langle \Phi_{ e } ( 1 ) r_{ 1 2 } \Phi_{ ( p_{ f } + 2 , \cdots ) } ( 2 ) \rangle  +   \langle \Phi_{ e } ( 1 ) r_{ 1 2 } \Phi_{ ( p_{ f }, q_{ f } + 2 , \cdots ) } ( 2 ) \rangle  \\ -2 \langle \Phi_{ ( p_{ e } + 1 , q_{ e } + 1 , \cdots )} ( 1 ) r_{ 1 2 } \Phi_{ ( p_{ f } + 1 , q_{ f } + 1 , \cdots ) } ( 2 )  \rangle  - \langle \Phi_{ e_{ + } } ( 1 ) r_{ 1 2 } \Phi_{ f_{ - } } ( 2 ) \rangle    \\    - \langle \Phi_{ e_{ - } } ( 1 ) r_{ 1 2 } \Phi_{ f_{ + } } ( 2 ) \rangle    \left. \right]  , \\   e_{ + } = ( p_{ e } , q_{ e } , \gamma_{ e } + 1 , \nu_{ e } + 1 , \alpha_{ e } , \beta_{ e } , m_{ e } + 1  ) , \\    e_{ + } = ( p_{ e } , q_{ e } , \gamma_{ e } + 1 , \nu_{ e } + 1 , \alpha_{ e } , \beta_{ e } , m_{ e } - 1  )                                                 \end{multline}       
 \begin{multline}  \langle r_{ 1 3 }^{ 2 } / r_{ 1 2 } \rangle = \int d \tau \Phi_{ a } ( 1 ) \Phi_{ s } ( 2 ) \Phi_{ t } ( 3 ) r_{ 1 3 }^{ 2 } / r_{ 1 2 }   =  \frac{ R^{ 2 } }{ 4 }   \left\{  \right.  \langle  \Phi_{ t } ( 3 ) \rangle  \\\times  \left[  \right.  \langle \Phi_{ ( p_{ a } + 2 , \cdots ) } ( 1 ) ( 1 / r_{ 1 2 } )  \Phi_{ s } ( 2 ) \rangle  +   \langle \Phi_{ ( p_{ a }  , q_{ a } + 2  \cdots ) } ( 1 ) ( 1 / r_{ 1 2 } )  \Phi_{ s } ( 2 ) \rangle   \\  - \langle \Phi_{ a }  ( 1 ) ( 1 / r_{ 1 2 } )  \Phi_{ s } ( 2 ) \rangle     \left.  \right]     + \langle \Phi_{ a } ( 1 ) ( 1 / r_{ 1 2 } ) \Phi_{ s } ( 2 ) \rangle  \\  \times \left[ \right. \langle \Phi_{  ( p_{ t} + 2 , \cdots ) }( 3 )  \rangle  +     \langle \Phi_{  ( p_{ t } , q_{ t} + 2 , \cdots ) }( 3 )  \rangle   -   \langle \Phi_{ t }( 3 )  \rangle   \left.   \right] \\ - \langle \Phi_{ t_{ - } } ( 3 ) \rangle  \langle \Phi_{ a_{ + } } ( 1 ) ( 1 / r_{ 1 2 } ) \Phi_{ s } ( 2 ) \rangle   -    \langle \Phi_{ t_{ + } } ( 3 ) \rangle  \langle \Phi_{ a_{ - } } ( 1 ) ( 1 / r_{ 1 2 } ) \Phi_{ s } ( 2 ) \rangle  \\ -2 \langle \Phi_{ ( p_{ t } + 1 , q_{ t } + 1 . \cdots ) } ( 3 ) \rangle \langle \Phi_{ ( p_{ a } + 1 , q_{ a } + 1 , \cdots ) } ( 1 ) ( 1 / r_{ 1 2 } ) \Phi_{ s } ( 2 ) \rangle \left. \right \} ,               \end{multline}                                         \begin{multline}  \langle 1 / r_{ 1 2 }^{ 3 } \rangle  =  \int d  \tau \Phi_{ a } ( 1 ) \Phi_{ b } ( 2 ) ( 1 / r_{ 1 2 }^{ 3 } )   =  \pi^{ 2 } R^{ 3 } \delta ( m_{ a } + m_{ b } ; 0 ) \\  \times \sum_{ l = 0 }^{ \infty }  \sum_{ m = | m_{ a } |  + 1  }^{ l } \frac{ 1 }{ 2 }  | ( - 1 )^{ m_{ a } + 1 } + ( - 1 )^{ m } |  m  ( 2 l + 1 )   \\ \times  Z_{ l }^{ m }    \int_{ 1 }^{ \infty } F_{ l }^{ m } ( z ) K_{ l , l , a }^{ m } ( z ) K_{ l , l, b }^{ m } ( z ) d z  ,    \end{multline}      
\begin{multline}     \langle 1 / r_{ 1 2 }^{ 2 } \rangle  = \int d \tau \Phi_{ a } ( 1 ) \Phi_{ b } ( 2 ) ( 1 / r_{ 1 2 }^{ 2 } )  =  \frac{ 1 }{ 2 }  \pi^{ 2 } R^{ 4 } \delta ( m_{ a } + m_{ b } ; 0 )  \\ \times  \sum_{ n = 0 }^{ \infty }  \sum_{ l = | m_{ a } | }^{ n } \left[ \frac{1 + ( - 1 )^{ l + m_{ a } } }{ 2 } \right]  \frac{ ( l !)^{ 2 } ( n - l ) ! ( n + 1 ) ( 2 l + 1 ) }{ ( n + l ) ! }   \\ \times \frac{ ( l - m_{ a } ) ! ( l + m_{ a } ) ! }{ \left[ (  \frac{ l - m_{ a } }{ 2 } ) !  \right]^{ 2 }    \left[ (  \frac{ l + m_{ a } }{ 2 } ) !  \right]^{ 2 }   }  \int_{ 1 }^{ \infty } \frac{ d z L_{ n - l, n - l, a ( l ) }^{ l + 1 } ( z )    L_{ n - l, n - l, b ( l ) }^{ l + 1 } ( z ) }{ ( z^{ 2 } - 1 )^{ l + 3 / 2 } \left[ C_{ n - l }^{ l + 1 } ( z ) \right]^{ 2 } } ,  \\            a ( l ) = ( p_{ a } , q_{ a } , \gamma_{ a } + l , \nu_{ a } + l , \alpha_{ a } , \beta_{ a } , m_{ a } ), \\  b( l ) = ( p_{ b } , q_{ b } , \gamma_{ b } + l , \nu_{ b } + l , \alpha_{ b } , \beta_{ b } , m_{ b } )  ,      \end{multline}        \begin{multline}  L^{ l }_{ n , n^{ \prime } , s } ( z ) = \int_{ 1 }^{ z }  \int_{ - 1 }^{ 1 } d \xi d \eta \xi^{ p_{ s }  } \eta^{ q_{ s } } ( \xi^{ 2 } - 1 )^{ \gamma_{ s } / 2 } ( 1 - \eta^{ 2 } )^{ \nu_{ s } / 2 }   \\  \times    e^{ - \alpha_{ s }  \xi } e^{ \beta_{ s } \eta } C_{ n }^{ l } ( \xi )  C_{ n^{ \prime } }^{ l } ( \eta )  ,                         \end{multline}           \begin{multline}   C_{ n }^{ l } { x } = \sum_{ j = 0 }^{ [ n / 2 ] }   \frac{ 2^{ n - 2 j } ( - 1 )^{ j } ( l + n - j - 1 ) !   \, x^{ n - 2 j } }{ j ! ( n - 2 j ) ! ( l - 1 ) ! }                                           \end{multline}     The upper limit of the sum over j is $ \frac{ 1 }{ 2 } n $ or $ \frac{ 1 }{ 2 } ( n - 1 ) $ , whichever is integral.  We $ \text{have}^{ 36 , 37 } $           \begin{multline}  \left( \frac{ 2 r_{ 1 2 } }{ R } \right)^{ -2 p } = \sum_{ n = 0 }^{ \infty } \sum_{ l = 0 }^{ n } d_{ n l } ( p ) \left[ ( 1 - \eta_{ 1 }^{ 2 } ) ( 1 - \eta_{ 2 }^{ 2 } ) ( \xi_{ 1 }^{ 2 } - 1 ) ( \xi_{ 2 }^{ 2 } - 1 ) \right]^{ 1 / 2 }    \\ \times D_{ n - l }^{ p + l } ( \xi_{ 1 > 2 } ) C_{ n - l }^{ p + l } ( \xi_{ 2 < 1 } ) C_{ n - l }^{ p + l } ( \eta_{ 1 } ) C_{ n- l }^{ p + l } ( \eta_{ 2 } ) C_{ l }^{ p - 1 / 2 } [ \cos ( \phi_{ 1 } - \phi_{ 2 } ) ] , \\  \quad  \quad \quad \quad   \quad   p > 0 , \quad  p \neq \frac{ 1 }{ 2 } \\  d_{ n l } ( p ) = \frac{ - 2^{ 2 l + 1 } \Gamma ( 2 p - 1 )  [  \Gamma ( p + l ) ]^{ 2 }  ( n - l ) ! ( n + p ) ( 2 p + 2 l - 1 )  }{ [ \Gamma ( p ) ]^{ 2 } \Gamma ( 2 p + n + l ) } ,  \\ D_{ m }^{ u } ( \xi ) = - C_{ m }^{ u } ( \xi ) \int_{ \xi }^{ \infty } \frac{ ( x^{ 2 } - 1 )^{ - u - 1 / 2 }   d   x  }{ [ C_{ m }^{ u } ( x ) ]^{ 2 } }  ,  \\  \frac{ d }{ d \xi } \left[ \frac{ D_{ m }^{ u } ( \xi ) }{  C_{ m }^{ u } ( \xi )     } \right]  =  \frac{ ( \xi^{ 2 } - 1 )^{ - u - 1 / 2 }  } { [ C_{ m }^{ u } ( \xi ) ]^{ 2 } }  ,     \end{multline}   \begin{multline}   \langle 1 / r_{ 1 2 } r_{ 1 3 } \rangle = \int d \tau \Phi_{ a } ( 1 ) \Phi_{ b } ( 2 ) \Phi_{ c } ( 3 )  1 / r_ { 1 2 } r_{ 1 3 }   =    \\    \frac{ 1 }{ 16 } \pi^{ 3 } R^{ 7 } \delta  ( m_{ a } + m_{ b } + m_{ c } ; 0 ) \delta ( \sigma ; | m_{ b } | ) \delta ( \sigma^{ \prime }  ; | m_{ c } | )   \left[ 1 + O_{ per } \binom{ b }{ c } \right] \\ \times \sum_{ l = \sigma }^{ \infty } \sum_{ j = \sigma^{ \prime } }^{ \infty } ( 2 l + 1 ) ( 2 j + 1 ) Z_{ l }^{ \sigma } Z_{ j }^{ \sigma^{ \prime } } \\ \times      \int_{ 1 }^{ \infty } d z N_{ l , j, a }^{ \sigma , \sigma^{ \prime } } ( z )  K_{ l , l , b }^{ \sigma } ( z ) \mathfrak{K}^{ \sigma^{ \prime } }_{ j , j , c } ( z ) F_{ l }^{ \sigma } ( z ) ,                                                        \end{multline}   \begin{center}   ACKNOWLEDGMENTS  \end{center}                                 This work was begun when the author was a graduate student at the University of California, Berkeley.  The author thanks Professor F. E. Harris for his willingness to discuss this problem and for checking some of the integrals.  The author thanks Professor B. Kirtman for discussions on the correlation problem, and Professor K. Street and Dr. A. Hebert for their encouragement.  This work was done under the auspices of the U. S. Atomic Energy Commission.  \\ This paper originally appeared in Physical Review A Vol. 3 Number 5 May 1971 page 1581.  The diagrams for the ten distinct integral types have been omitted.  The title has been simplified.  Typos in the equations have been corrected in Equations 4, 7, 11, 13, 13, 22, 22, 35, 35, 39, 39 , 44 , 49 and the unnumbered equation for $ \Psi_{ tot } $ before [Eq. (53)] .                 \\    \small{ 1.  C.L.Pekeris, Phys. Rev. \underline{115},1217(1959) } \\             \small{ 2. E.A.Burke, Phys. Rev. \underline{130},187(1963) }  \\                \small{ 3. W.Kolos and C.C.J.Roothaan, Rev.Mod.Phys. \underline{32},205(1960) } \\  \small{ 4. L.Szasz, Z.Naturforsch. \underline{15a},909(1960) } \\            \small{ 5. F.E.Harris,J.Chem.Phys. \underline{32},3(1960) } \\                   \small{ 6. M.Kotani,\textit{Tables of Molecular Integrals}(Maruzen,Tokyo,1955),Chap. 1 } \\                                                                   \small{ 7. P.M.Morse and H.Feschbach, \textit{Methods of Theoretical Physics} (McGraw-Hill, New York,1953), Vol. 1 p. 655 } \\                                   \small{ 8. I.Shavitt,in \textit{Methods In Computational Physics}(edited by B.J.Alder, S.Fernbach and M.Rotenberg(Academic,New York,1963), Vol.2, p.1 } \\    \small{ 9. M.P.Barnett, in Ref. 8, p. 153 } \\                                   \small{10. C.Berge,\textit{Theorie des Graphes et Ses Applications},(Dunod,Paris,1963), p.153 } \\                                                              \small{11. G.Polya, AM.Math.Monthly \underline{63},689(1956) } \\               \small{12. F.E.Neumann,J.Reine Angew.Math (Crelle) \underline{37},21(1848) ;     \textit{Vorlesungen Uber die Theorie des Potentials und der Kugelfunktionen}   (Teubner,Leipzig,1878)  Chap.13 }  \\                                           \small{13. E.Jahnke and F.Emde,\textit{Tables of Functions}(Dover,New York,1945) }    \\                                                                           \small{14. \textit{Higher Transcendental Functions} edited by A.Erdelyi, W.Magnus, F.Oberhettinger and F.G.Tricomi(McGraw-Hill,New York,1954),Vol.2 } \\    \small{15.C.S.Meijer, Proc. Nederl. Akad. Wetensch. \underline{42},930(1939)} \\ \small{16. G.Frobenius,J.Reine Angew. Math(Crelle) \underline{73},1(1871) } \\    \small{17. E.W.Hobson,\textit{Spherical and Ellipsoidal Harmonics}(Cambridge U.P.,London,1931)}   \\                                                         \small{18. L.W.Thome,J.Reine Angew.Math(Crelle)\underline{66},337(1866)} \\     \small{19. W.E.Byerly,\textit{An Elementary Treatise on Fourier's Series and Spherical, Cylindrical and Ellipsoidal Harmonics, with Applications to Problems in Mathematical Physics},(Ginn,Boston,1893) }  \\                                  \small{20. J.A.Gaunt,Phil.Trans.Roy.Soc. London \underline{A228},151(1928) } \\ \small{21. C.G.Darwin, Proc.Roy.Soc.London \underline{A118},654(1928) } \\      \small{22. J.C.Adams,Proc.Roy.Soc. London \underline{27},63(1878) } \\          \small{23. J.D.Talman,\textit{Special Functions Based on Lectures by E.P.Wigner} (Benjamin,New York, 1968) }  \\                                                \small{24. F.E.Harris and H.H.Michels, in \textit{Advances In Chemical Physics}, edited by I.Prigogine (Interscience, New York, 1967),Vol.13, p. 205 } \\       \small{25. K.Rudenberg,J.Chem.Phys.\underline{22},765(1954) } \\                 \small{26. K.Rudenberg,J.Chem.Phys.\underline{19},1459(1951) } \\              \small{27. F.E.Harris (unpublished) } \\                                        \small{28. Reference 24 Appendix B } \\                                          \small{29. U.Fano and G.Racah, \textit{Irreducible Tensorial Sets}.(Academic,New York, 1959), p.36 } \\                                                         \small{30. E.U.Condon and G.Shortley, \textit{The Theory of Atomic Spectra}, (Cambridge U.P.,London,1935) } \\                                                  \small{31. N.Ja.Vilenkin,\textit{Special Functions and the Theory of Group Representations},(American Mathematical Society,Providence,R.I.,1968) } \\            \small{32. E.P.Wigner (unpublished) } \\                                         \small{33. G.Racah, Phys. Rev. \underline{62},438(1942) } \\                    \small{34. B.R.Judd,\textit{Operator Techniques in Atomic Spectroscopy},(McGraw-Hill,New York, 1963) } \\                                                    \small{35. M.Tinkham,\textit{Group Theory and Quantum Mechanics}(McGraw-Hill, New York,1964) } \\                                                               \small{36. L.Wolniewicz, Acta.Phys.Polon.Suppl.\underline{22},3(1962) } \\        \small{37. W.Kolos and L.Wolniewicz,Acta.Phys.Polon.\underline{20},129(1961) }                                                                                   \end{document}